\documentclass[9pt, sigconf]{acmart}
\usepackage{times}
\usepackage{graphicx}
\usepackage{balance}  
\usepackage[ruled,vlined,linesnumbered]{algorithm2e}
\usepackage{threeparttable}
\usepackage{multirow}
\usepackage{graphicx}
\usepackage{tikz}
\usepackage{enumitem}
\setlist{nolistsep}
\usepackage{makecell}
\usepackage[labelfont=bf,skip=0pt]{caption,subfig}
\usepackage{hyperref}
\usepackage{xcolor}
\usepackage{dsfont}
\usepackage{euscript}
\renewcommand{\Pr}[1]{\ensuremath{\mathbf{Pr}\!\left[#1\right]}}
\newcommand{\Var}{\mathrm{Var}}
\newcommand{\E}{\mathbb{E}}
\newcommand{\card}[1]{\ensuremath{\lvert#1\rvert}}
\def\mparagraph#1{\par\smallskip\noindent\textbf{#1.}\quad}
\newcommand{\Times}{\mathbb{T}}
\newcommand{\States}{\mathcal{X}}
\newcommand{\State}[1]{X_{#1}}
\newcommand{\SimStep}{\mathfrak{g}}

\DeclareMathOperator*{\argmin}{arg\,min}

\newtheorem{lemma}{Lemma}
\newtheorem{proposition}[lemma]{Proposition}

\definecolor{reva}{rgb}{0,0,0}
\definecolor{revb}{rgb}{0,0,0}
\definecolor{revc}{rgb}{0,0,0}
\definecolor{revmeta}{rgb}{0,0,0}
\definecolor{default}{rgb}{0,0,0}

\newcommand{\ansa}[1]{{{\color{reva}{#1}}}}
\newcommand{\ansb}[1]{{{\color{revb}{#1}}}}
\newcommand{\ansc}[1]{{{\color{revc}{#1}}}}

\AtBeginDocument{%
  \providecommand\BibTeX{{%
    \normalfont B\kern-0.5em{\scshape i\kern-0.25em b}\kern-0.8em\TeX}}}

\copyrightyear{2021}
\acmYear{2021}
\setcopyright{acmcopyright}
\acmConference[SIGMOD '21] {Proceedings of the 2021 International Conference on Management of Data}{June 20--25, 2021}{Virtual Event, China}
\acmBooktitle{Proceedings of the 2021 International Conference on Management of Data (SIGMOD '21), June 20--25, 2021, Virtual Event, China}
\acmPrice{15.00}
\acmISBN{978-1-4503-8343-1/21/06}
\acmDOI{10.1145/XXXXXX.XXXXXX}
\begin{document}

\title{Efficiently Answering Durability Prediction Queries\\ (Technical Report Version)}

\author{Junyang Gao}
\authornote{Most of the work was conducted when author was at Duke University.}
\email{jygao@google.com}
\affiliation{%
  \institution{Google Inc.}
}
\author{Yifan Xu}
\email{xuyifa@amazon.com}
\affiliation{%
  \institution{Amazon.com}
}
\author{Pankaj K. Agarwal}
\email{pankaj@cs.duke.edu}
\affiliation{%
  \institution{Duke University}
}
\author{Jun Yang}
\email{junyang@cs.duke.edu}
\affiliation{%
  \institution{Duke University}
}

\begin{abstract}
We consider a class of queries called \emph{durability prediction queries} that arise commonly in predictive analytics, where we use a given predictive model to answer questions about possible futures to inform our decisions.
Examples of durability prediction queries include ``what is the probability that this financial product will keep losing money over the next 12 quarters before turning in any profit?'' and ``what is the chance for our proposed server cluster to fail the required service-level agreement before its term ends?''
We devise a general method called \emph{Multi-Level Splitting Sampling (MLSS)} that can efficiently handle complex queries and complex models---including those involving black-box functions---as long as the models allow us to simulate possible futures step by step.
Our method addresses the inefficiency of standard Monte Carlo (MC) methods by applying the idea of \emph{importance splitting} to let one ``promising'' sample path prefix generate multiple ``offspring'' paths, thereby directing simulation efforts toward more promising paths.
We propose practical techniques for designing splitting strategies, freeing users from manual tuning.
Experiments show that our approach is able to achieve unbiased estimates and the same error guarantees as standard MC while offering an order-of-magnitude cost reduction.
\end{abstract}
\settopmatter{printfolios=true, authorsperrow=4}
\maketitle
\setcounter{page}{1}
\section{Introduction}
\label{sec:intro}

Increasingly, we rely on \emph{predictive analytics} to inform decision making.
Typically, we build a model to predict the future, using historical data and expert domain knowledge.
Then, using this model, we can ask questions about possible futures to inform our decisions.
A common type of such questions are what we call \emph{durability prediction queries}, which predict how likely is it that a condition will remain over a given duration into the future.
For example, business analysts ask durability prediction queries for financial risk assessments: ``what is the probability that this financial product will keep losing money over the next 12 quarters before turning in any profit?'' or ``how likely is it that our client will not have a default on the mortgage loan in the next five years?''
As another example, engineers ask durability prediction queries about reliability: ``what is the probability that a self-driving car makes a serious misjudgement within its warranty period?'' or ``what is the chance for our proposed server cluster to fail the required service-level agreement before its term ends?''
In this paper, we consider the problem of answering durability prediction queries given a predictive model of the future. 

Durability prediction queries are challenging for several reasons.
In contrast to queries over historical data, these queries must deal with uncertainties about the possible futures.
Moreover, temporal dependence broadly exists in temporal data: across examples such as financial markets, customer behaviors, or system workloads, the state of the world at the present time often depends strongly on the recent past.
This observation renders inapplicable much of the existing work on query processing over probabilistic databases~\cite{re2007efficient, dalvi2009probabilistic, soliman2007top,hua2008efficiently,yi2008efficient, hua2011ranking, ge2009top}, where uncertainty in data is assumed to be independent across objects (e.g., attribute or tuple values).

\ansc{
A second challenge stems from the growing popularity of complex predictive models.
Oftentimes, due to cost and data privacy considerations, we do not have the luxury of building a custom model directly just to answer one specific durability prediction query; instead, we would be given a general model with which we can simulate possible future states of data and use them to answer various queries.
This paper does not prescribe how to come up with such a model; we assume it is given to us and focus on how to use it to answer durability prediction queries efficiently.
}
However, we do want to support a wide gamut of models---be it a traditional stochastic processes with good analytical properties or a black-box model that powers its predictions by deep neural networks.
While it is possible to derive analytical answers to durability prediction queries for simple queries and simple models on a case-by-case basis (e.g., when the value of an auto-regressive process~\cite{shumway2017time} hits a particular threshold), our goal is to derive a general-purpose procedure that works for any query and for any model, provided that the model allows us to simulate possible futures step by step.
For example, the simulation model can use a recurrent neural network~\cite{rumelhart1985learning, jordan1997serial, hochreiter1997long} to predict prices for a collection of stocks for the next hour using their prices during the past 36 hours, and the query can ask for the probability that a given stock's P/E ratio will rank among the top 10 by the end of the week.
In general, we will need to resort to Monte Carlo (MC) techniques~\cite{binder1993monte} to generate multiple ``sample paths'' (each corresponding to one possible sequence of future states) and evaluate the query condition on them, in order to derive an estimate to the query answer.

A third challenge, however, stems from the serious inefficiency of the standard MC technique of simple random sampling.
It suffers greatly when the answer probability is small.
In fact, as previous studies have pointed out~\cite{l2006splitting, o2018scalable}, the \emph{relative error} of standard MC increases to infinity as the underlying probability approaches $0$, therefore requiring a prohibitively expensive number of simulations in order to achieve acceptable error.
\ansb{
Unfortunately, in many practical use cases of durability prediction queries such as the examples above, people are interested in looking for robust and consistent behaviors over time, which naturally leads to small answer probabilities.
}
Hence, the drawbacks of simple random sampling are further amplified by durability prediction queries.

To address these challenges, we propose an alternative approach to answer durability prediction queries that preserves the generality and simplicity of standard MC, but provides significant efficiency improvement.
The key insight is that not all sample paths are equally promising.
Instead of generating each sample path independently from the very start (as in simple random sampling), we can gauge, from where a path has been so far, how close it is to hitting the condition of interest.
We would then ``split'' a promising partial path into multiple ``offsprings,'' by continuing multiple simulations from this same partial path.
With this approach, we effectively direct more simulation efforts towards those more promising simulation paths.
Note that we achieve this goal simply by choosing when and where (during sampling) to invoke the given per-step simulation procedure (which can be an arbitrarily complex black-box) without changing its internals, which makes this approach very general and practical.

Our main technical contributions are as follows:
\begin{itemize}[leftmargin=*]
\item We formalize the notion of \emph{durability prediction queries} given a predictive model with a step-by-step simulation procedure.
The generality of our novel problem formulation and solutions means that they are widely applicable, even for complex models and complex queries that are increasingly common in practice.
\item 
\ansb{
We propose \emph{Multi-Level Splitting Sampling (MLSS)} as a general method for answering durability prediction queries.
This method applies the idea of \emph{importance splitting}~\cite{garvels2000splitting}, which has been well-studied in statistic community~\cite{l2006splitting,cerou2007adaptive,villen1994restart}.
However, the original idea has limited applicability because of several strong assumptions on the underlying stochastic process,
and it will produce incorrect estimates when applied blindly to our problem.
Our approach drops many unrealistic assumptions and is generally applicable on a larger class of processes;
it still provides significant efficiency improvement with provably unbiased estimates.}
%
\item Practical application of MLSS requires designing ``levels'' that correspond to ``progress milestones'' where we split a sample path upon reaching them.
We further propose an adaptive greedy strategy that automatically and incrementally searches for a good level design, thereby freeing users from manual tuning.
The strategy incurs low overhead and obtains good level design in practice.
%
\end{itemize}

\section{Preliminaries}\label{sec:ps}
\subsection{Problem Formulation}\label{sec:formulation}
\mparagraph{Stochastic Process and Simulation Model}
Consider a discrete time domain of interest $\Times = \{0,1,2,3,\dots\}$ and a discrete-time stochastic process $\{\State{t}\}_{t \in \Times}$ with state space $\States$, where $\State{t} \in \States$ is a random variable of state at time $t$.
We are given an \emph{initial state} $x_0$ and a \emph{step-wise simulation procedure} $\SimStep$ that simulates the process forward step by step: given previous states $x_{<t} = \{x_0, \dots, x_{t-1}\}$ up to time $t-1$, $\SimStep(x_{<t}, t)$ returns (randomly) the state $x_t$ at time $t$ ($x_t$ is the observed value of $X_t$).
Starting from $x_0$, we can generate a \emph{sample path} $\{ x_0, x_1, x_2, \ldots \}$ of arbitrary length for the process by repeatedly invoking $\SimStep$.
Multiple sample paths can be generated simply by restarting the sequence of invocations.

This definition covers a wide range of generative models used in practice for temporal data.
It is worth noting that we place no restriction on how complex the state space $\States$ and the step-wise simulation procedure $\SimStep$ are.
We highlight some examples:

(1) \textit{Auto-regressive or AR($m$) model}: 
Here the simulation procedure $\SimStep$ draws the value $v_t$ at time $t$ according to values of the last $m$ time steps, $\{v_{t-1}, v_{t-2}, \ldots, v_{t-m}\}$, by $\sum_{i=1}^m \phi_i v_{t-i} + \epsilon_t$, where $\phi_i$'s are model parameters and $\epsilon_t$'s are random errors.

(2) \textit{Time-homogeneous (discrete-time) Markov chains}: Here $\SimStep$ generates the state at time $t$ according to a given probability distribution $\Pr{\State{t} \mid \State{t-1}}$ independent of $t$ and conditionally independent of all states prior to the last.

(3) \textit{Black-box models}: The popularity of deep learning in recent years has given rise of highly complex models that are difficult to reason with analytically.
Consider, for example, a model that captures the relationship between successive states using a recurrent neural network (RNN).
Here, $\SimStep$ generates the value $v_t$ for time $t$ according to $v_t \sim o(g(h_{t-1}, v_{t-1}; \theta); \theta)$, where $h_{t-1}$ denotes the state of the hidden layer(s) at time $t-1$, $o(\cdot)$ and $g(\cdot)$ are activation functions, and $\theta$ denotes (time-invariant) model parameters; $\SimStep$ further updates $h_{t-1}$ to $h_t$.
The state at time $t$ hence includes both $v_t$ and $h_t$.
\mparagraph{Durability Prediction Queries}\hspace{-1em}
Given a stochastic process $\{\State{t}\}_{t \in \Times}$ governed by $\SimStep$ with initial state $\State{0}$, let $q: \States \to \{0,1\}$ be a user-specified Boolean \emph{query function} that returns $1$ if a given state satisfies a condition of interest (and $0$ otherwise).
A \emph{durability prediction query} (or \emph{durability query} for short) $Q(q,s)$
returns the probability that the process ever reaches any state for which $q$ returns $1$ by the end of the prescribed time horizon $s \in \Times$.
Formally, 
$Q(q,s) = \Pr{ \smash{\bigvee_{1 \le t \le s}} q(\State{t}) = 1}$.
Alternatively, consider the time $T$ before the process first ``hits'' (meets) the condition of interest; $T$ is an random variable, and the durability query $Q(q,s)$ returns $\Pr{T \le s}$, i.e., the probability that the hitting time is within the prescribed threshold $s$.
\ansb{As mentioned above, $Q(q,s)$ tends to be small probabilities in real-life applications.}

To illustrate, suppose we use a RNN-based model described earlier to predict the price and earning for $n$ stocks.
The state at time $t$ consists of $h_t$ (of the hidden layers) as well as a vector $v_t = \smash{\langle p^{(1)}_t, e^{(1)}_t, \dots p^{(n)}_t, e^{(n)}_t \rangle}$, where $\smash{p^{(i)}_t}$ and $\smash{e^{(i)}_t}$ are the price and earning for stock $i$ at time $t$.
For a durability query $Q(q,s)$ concerning the probability that stock $i$ can break into top $10$ by time $s$ in terms of P/E ratio, the query function $q$ would access the vector of prices and earnings in the state, compute all P/E ratios, and check if $i$'s rank is within $10$.
$Q(q,s)$ is the probability that a random sample path reaches a state for which $q$ evaluates to $1$ within time $s$.

For simple models and simple query functions (e.g., the condition of interest is whether an AR($m$) process exceeds a given value), we can in fact compute durability prediction queries analytically and exactly.
In general, with complex models or complex query functions, computing durability prediction queries becomes exceedingly difficult.
Especially when the model itself is complex, we have to resort to Monte Carlo simulations using $\SimStep$.
Let $\tau = Q(q,s)$ denote the exact answer to the query.
Instead of returning $\tau$, our goal is to devise an algorithm that can produce an unbiased estimate $\hat\tau$ of $\tau$ together with some statistical quality guarantee (e.g., confidence interval or estimator variance).
We measure the cost of the algorithm by the total number of invocations of $\SimStep$.
In practice, the user can specify a cost budget, and our algorithm will produce a final estimate with quality guarantee when the budget runs out.
Alternatively, the user can specify a target level of quality guarantee, and our algorithm will run until the target guarantee is reached.
In this paper, we are interested in achieving better guarantees given a fixed budget, or achieving the target guarantee with lower costs.

\subsection{Background and Other Approaches}\label{sec:background}


Durability prediction queries are deeply connected to a classic problem in statistics called \emph{first-hitting time} or \emph{first-passage time} in stochastic system~\cite{cox1977theory,redner2001guide, whitmore1986first}.
Similar problems related to first-hitting time are also independently studied in very diverse fields, from economics~\cite{shiryaev1999essentials} to ecology~\cite{fauchald2003using}.
We briefly introduce several existing approaches for durability prediction queries (or the first-hitting time problem) here,
and lay the foundation for later sections.

\mparagraph{Analytical Solution}
As mentioned earlier, there exist analytical solutions for some simple stochastic processes~\cite{grimmett2001probability}, e.g., Random Walks, AR($m$) model, to name but a few.
However, real applications often require more complex structures.
For instance, Compound-Poisson process is a well-known stochastic model for risk theory in financial worlds.
In~\cite{xu2012first}, authors derived an analytical solution for such stochastic processes.
However, the exact solution itself is very complicated, involving multiple integrals that still require numerical approximations.
In general, the analytical solution to first-hitting problem is model-specific, may not exist for most applications, and hence cannot be directly used for durability query processing.

\mparagraph{Simple Random Sampling (SRS)}
Monte Carlo simulations is the most general approach for answering durability prediction queries.
SRS is the standard Monte Carlo technique.
To answer durability query $Q(q, s)$ with query function $q$ and prescribed time threshold $s$, we randomly simulate $n$ independent sample paths according to the procedure $\SimStep$.
For each sample path $SP_i = \{x_0,x_1, \dots\}$, we define a label function indicating whether the simulated path satisfies the query condition:
\[
l(SP_i) = 
\begin{cases}
        1, ~~~\bigvee_{1 \le t \le s} q(x_t) = 1,\\
        0, ~~~\text{otherwise}.
    \end{cases}
\]
Then, an unbiased estimator of SRS is $\hat{\tau}_{srs} = \frac{\sum_{i=1}^n l(SP_i)}{n}$, 
with estimated variance $\widehat{\Var}(\hat{\tau}_{srs}) = \frac{\hat{\tau}_{srs}(1-\hat{\tau}_{srs})}{n}$.

We use SRS as the main baseline solution throughout the paper.
The major drawback of SRS is its ``blind search'' nature---it randomly simulates sample paths and just hopes that they could reach the target.
For durability prediction queries with small ground truth answer $\tau$, SRS would waste significantly much simulation effort on those sample paths that do not ever satisfy the query condition.

\mparagraph{Importance Sampling (IS)}
Importance sampling is one of the most popular variance reduction techniques for Monte Carlo simulations.
It is a special case of biased sampling, where sampling distribution systematically differs from the underlying distribution in order to obtain more precise estimate using fewer samples.
Let us use the following concrete example for illustration.
Consider an AR(1) model, where simulation procedure $\SimStep$ draws the value $v_t$ according to $\displaystyle \phi_1 v_{t-1} + \epsilon_t$.
Here, $\phi_1$ is a constant parameter and $\epsilon_t$ is independent Gaussian noise; i.e., $\epsilon_t \sim N(0, \sigma)$ for $t \in \Times$.
Given time threshold $s$, the random variable of interest $l(SP)$ has probability density $g(l) \sim \prod_{i=1}^s N(0, \sigma)$.
IS draws samples from an instrumental distribution $\omega$, and an unbiased estimator is $\displaystyle \hat{\tau}_{is} = \frac{1}{n} \sum_{i=1}^n\frac{g(l(SP_i))}{\omega(l(SP_i))}l(SP_i)$.
Choosing a good instrumental distribution $\omega$ is critical for the success of IS. 
An iterative approach called \emph{Cross-Entropy (CE)}~\cite{de2005tutorial,rubinstein1997optimization} is widely used for importance sampling optimization.
However, IS typically requires a priori knowledge about the model, e.g., model parameters or state transition probabilities.
This requirement can be impractical for some complex temporal processes, not to mention black-box models that we consider in this paper.

\ansc{
\mparagraph{Learning Durability Directly}
Instead of answering durability prediction queries from using a simulation model,
one could also learn predictive models that \emph{directly} answer durability prediction queries.
However, a general simulation model as we considered in this paper is often preferred based on the following considerations: 
(1) In some cases, we do not always have the luxury of training custom models from real data points just to answer queries. It is costly or even unethical to collect training data for such purposes, e.g., autonomous driving car testing. 
(2) Building an one-off model to answer durability prediction queries (with different query conditions and different parameters) would quickly become infeasible because each type of durability prediction implies a different modeling exercise.  
In contrast, a general simulation model can be conveniently reused for answering a variety of queries using our technique without requiring extra data collections or modeling expertise. 
Moreover, a nice byproduct of utilizing simulation models is that we also produce a set of concrete sample paths alongside the point estimate and confidence interval. Users can look into these “possible worlds” to get a better understanding of query answers. 
Compared to the relatively opaque direct models that output only the final durability prediction, this approach provides more interpretability and credibility.
We acknowledge the difficulty of having good simulation functions, but we note that for many domains, e.g., finances, autonomous vehicles, etc., the use of such general simulation models are well-established and common.}

\mparagraph{Remarks}
All solutions reviewed above have limitations: analytical solutions and IS are not generally applicable to durability prediction queries; SRS can be very inefficient; \ansc{and the alternative of learning custom models to predict durability directly may be infeasible for practical (ethical or cost) reasons.}
In the ensuing sections, we introduce a novel approach for answering durability prediction queries from simulation models that achieves the generality of SRS as well as the efficiency of IS.

\section{Simple MLSS}\label{sec:mlss}

\begin{table}\small
    \centering
    \caption{Notations}
    \label{tab:notation}
    \tabcolsep=0.1cm
    \begin{tabular}{|c|c|}
    \hline
    $s$ & Prescribed time horizon of the durability query. \\\hline
    $r$ & Splitting ratio (or branching factor). \\\hline
    $m$ & Number of levels. \\\hline
    \ansa{$\beta_i/\beta$} & \ansa{The partition boundary for the $i$-th level / target value.} \\\hline
    $L_i$ & The $i$-th level. $L_m$ is the target level. \\\hline
    $N_i$ & \makecell{Number of first-time entrance state into $L_i$. \\ $N_0$ is the number of root paths; $N_m$ is the number of hits to the target.} \\\hline
    $p_i$ & \makecell{(conditional) Level advancement probability from level $L_{i-1}$ to $L_{i}$.}
    \\\hline
\end{tabular}
\vspace{-0.5em}
\end{table}

Since generating too many paths that do not meet the query condition can be a waste of simulation cost, it is natural to design a sampling procedure that more frequently produces paths that reach the target.
To this end we apply the idea of \emph{splitting}~\cite{garvels2000splitting}.
The intuition is to encourage further explorations of paths that are more likely to hit the condition of interest by splitting them into multiple offsprings when they reach particular ``milestones'' (see Figure \ref{fig:mlss-example} for illustration).
Such treatment is analogous to the notion of sample weight/importance in importance sampling, but without explicitly tweaking the underlying step-wise simulation procedure $\SimStep$.

\ansb{
Traditional applications of the splitting idea are mostly concerned with much simpler settings (e.g., process with strong Markov properties).
To apply splitting to our setting of durability prediction queries, we need to introduce the concept of \emph{value functions}.
}

\mparagraph{Value Functions}
We first capture how promising a path prefix is using a \ansb{heuristic} \emph{value function} $f(x_t): \States \times \Times \to (0,1]$. The closer $f(x_t)$ is to $1$, the more likely that the process will reach the query condition given the current state.
We further require that $f(x_t)=1$ if and only if $q(x_t)=1$.


\ansa{
As outlined in the definition of \emph{levels} below, $f$ guides when to split a path, thus a properly defined $f$ leads to more efficient simulation efforts. It is worth noting that the unbiasedness of our estimator in this section does \emph{not} depend on $f$; only its efficiency does.

The best choice of $f$ is problem-specific as it depends on both the simulation model and the query. In practical applications, the query condition often takes the form  $z(x_t) \ge \beta$, where $z: \States \rightarrow \mathbb{R}$ is a real-valued evaluation of a state and $\beta$ is a user-specified value threshold, and $z(x_t)$ has a higher chance to hit the boundary when $x_t$ is closer to it. Thus a reasonable value function would be $f(x_t) = \min\{z(x_t)/\beta, 1\}$. More sophisticated designs of value functions are certainly possible, but are beyond the scope of this paper.}

\mparagraph{Levels}
With the help of the value function $f$, we can now introduce the notion of \emph{levels} to capture multiple intermediate ``milestones'' a sample path can reach before meeting the query condition.
We partition $[0,1]$, the range of value function $f$, into $m+1$ disjoint \emph{levels} (intervals) with boundaries $0 = \beta_0 < \beta_1 < \cdots < \beta_m = 1$, where $L_i = [\beta_i, \beta_{i+1})$ for $0 \le i \le m-1$ are the first $m$ levels, and the degenerated interval $L_m = [1,1]$ is the last level.
Let $T_i(SP) = \inf\{t \ge 0 \mid f(x_t) \in L_i\}$ be the first time that a sample path $SP:\{x_t\}_{t\ge 1}$ enters level $L_i$.
Let $\Xi_i = \{SP \mid T_i(SP) \le s\}$ denote the event that the process enters $L_i$ before time threshold $s$; i.e., the set of all possible simulated paths that enter $L_i$ before $s$.

\subsection{s-MLSS Sampler and Estimator}
\label{sec:mlp}
\label{sec:sample-and-estimate}

We are now ready to describe a \emph{simple} version of \emph{Multi-Level Splitting Sampling}, \emph{s-MLSS}.
Frequently used notations are summarized in Table~\ref{tab:notation}.
As noted earlier, the idea of \emph{splitting} can be traced back to 1951~\cite{kahn1951estimation} and has been used by several authors in statistic community~\cite{garvels2000splitting}.
The interested readers can refer to \cite{garvels2000splitting} for a more comprehensive introduction.
\ansb {
However, prior studies mostly focused on stochastic processes with strong Markov property.
Here we introduce s-MLSS for more general simulation models in the context of durability prediction queries.
Nonetheless, s-MLSS still inherits a critical assumption from existing literature:
}

%
\noindent\textbf{(No Level-Skipping)} Any sample path generated from the stochastic procedure $\SimStep$ has to enter $L_i$ before it enters $L_{i+1}$, for every $i\leq m$.
%

\ansb{
This assumption is rather restrictive and does not hold in general.
It is possible (especially in practice with discrete time) for a sample path's value to jump, between two consecutive time instants, from a level to a higher one, crossing multiple levels in between.
We show with experiments in Section~\ref{sec:expr} that ignoring this assumption and blindly applying s-MLSS will in practice lead to incorrect answers.
In Section~\ref{sec:extension}, we instead see how \emph{g-MLSS}, our general version of MLSS, lifts this assumption and correctly handles the general case.
}
Nonetheless, s-MLSS serves as a good starting point for our exposition.

As a result of the no-level skipping assumption, we have the following containment relation: 
\begin{equation}\label{eq:mlss-assumption}
    \Xi_m \subset \Xi_{m-1} \subset \cdots \subset \Xi_1 \subset \Xi_0.
\end{equation}
With (\ref{eq:mlss-assumption}) and the chain rule for probability we decompose the target probability $\tau$ as
\begin{equation}\label{eq:mlss-truth}
\begin{split}
\resizebox{.9\hsize}{!}{
    $\tau = \Pr{\Xi_m} 
    = \Pr{\Xi_m \mid \Xi_{m-1}}\cdots\Pr{\Xi_1\mid \Xi_0}\Pr{\Xi_0} = \prod_{i=1}^m p_i$
},
\end{split}
\end{equation}
where  $p_i = \Pr{\Xi_i \mid \Xi_{i-1}}$ is referred to as the \emph{level advancement probability}.
Next we show the s-MLSS sampling approach that estimates $p_i$'s and in turn $\tau$.

\mparagraph{s-MLSS Sampler}
In a nutshell, s-MLSS works in rounds of stages, estimating the decomposed probability $p_i$ separately between consecutive levels.
For each level $L_i$, we maintain a counter $N_i$ denoting the number of sample paths that enter $L_i$ for the first time.
In the first stage, we start the simulation of a path from the initial level $L_0$, which we refer to as the \emph{root path}, and increment the counter $N_0$ by 1.
We continue the simulation up to time $s$:
\begin{enumerate}[leftmargin=*]
    \item If the sample path does not enter the next level $L_1$, we stop and start a new round of simulation for the next root path.
    \item Otherwise, we increment the counter $N_1$ by 1 and split the root path into $r$ independent copies at the first time it enters $L_1$, where $r$ is a constant called \emph{splitting ratio}.
    Assume the hitting time is $t$, we define the state $\mathcal{X}_{t}$ of sample path as the \emph{entrance state} to $L_1$.
    All splitting copies from the original path will use the same entrance state $\mathcal{X}_{t}$ as starting point for future simulations.
\end{enumerate}
Then in the next stage, for each of the splitting offspring of the root path, we recursively follow the similar procedure as described above: simulate the path up to time $s$; if it reaches the next level, increment the counter of that level, split and repeat;
If not, finish the simulation at time $s$.
The simulation of a root path stops when we finish the simulations of all its splitting offspring---either enters the target level $L_m$ or runs until the time $s$.

\begin{figure}
    \centering
    \includegraphics[width=0.2\textwidth]{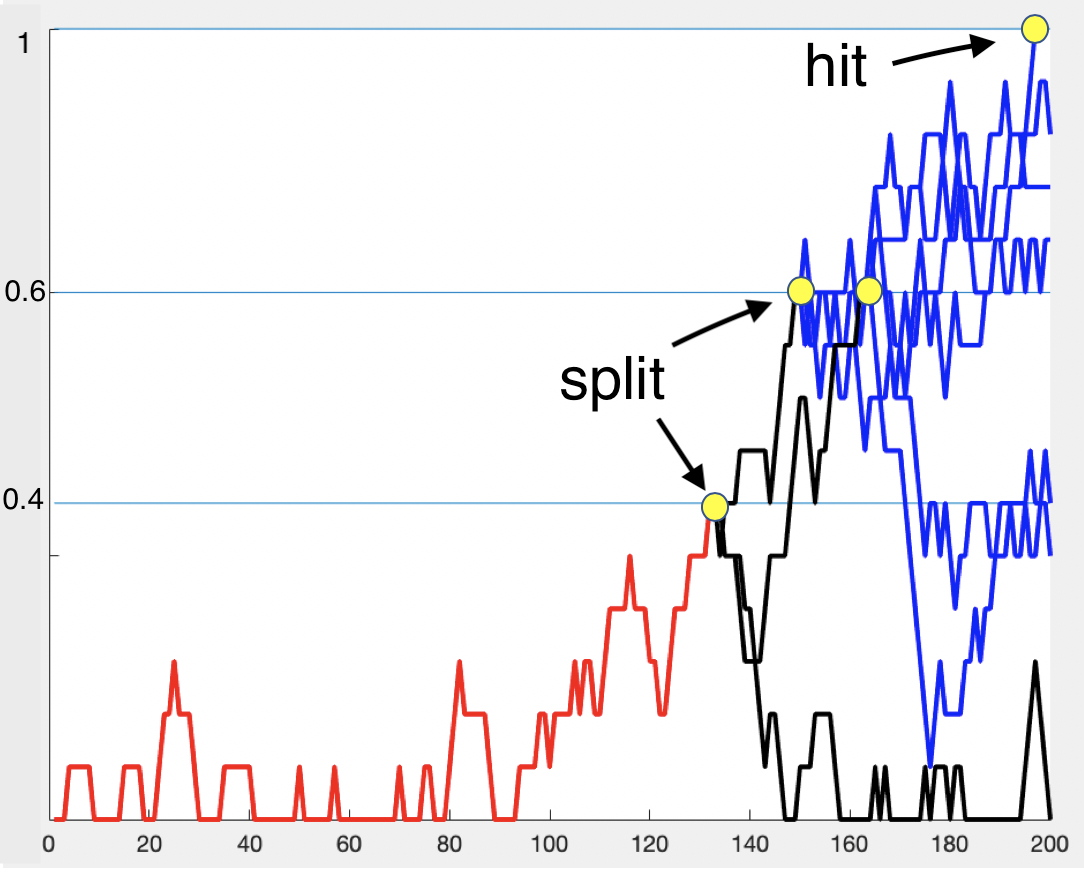}
    \vspace*{0.5ex}
    \caption{Simulations of a root path using MLSS with splitting ratio $r=3$. \mdseries Horizontal axis is time, with horizon $s=200$; vertical axis shows the result of the value function.}
    \label{fig:mlss-example}
\end{figure}

Figure~\ref{fig:mlss-example} illustrates a concrete example of the simulations of one root path.
Here we have $s=200$, levels $L_0=[0,0.4),L_1=[0.4,0.67),L_2=[0.67,1),L_3=[1,1]$, and splitting ratio $r=3$.
The root path (red line) starts from $L_0$ and enters $L_1$ at timestamp 133.
Then it splits into 3 copies (black lines) and continues the simulations forward.
Two out of the three splitting paths (from $L_1$) enter $L_2$ and each of them further splits into three more copies (blue lines), respectively.
Finally, one out of the six splits (from $L_2$) enters the target level $L_3$.
All other copies (that do not have the chance to split) run till time $s$ and stop.
Following the above procedure, assume we sample and simulate $N_0$ root paths until the stopping criteria is met (i.e., the simulation budget runs out or the estimate achieves the target quality guarantee).

\mparagraph{s-MLSS Estimator}
Using the counters we have maintained through MLSS for each level, we have $\smash{\hat{p}_1 = \frac{N_1}{N_0}, \hat{p}_2 = \frac{N_2}{rN_1}, \dots,}$ $\smash{\hat{p}_m = \frac{N_m}{rN_{m-1}}}$.
The estimator for MLSS is 
\begin{equation}\label{eq:mlss-est}
    \hat{\tau}_{mlss} = \prod_{i=1}^{m} \hat{p_i} = \frac{N_1}{N_0}\frac{N_2}{rN_1}\cdots \frac{N_m}{rN_{m-1}} = \frac{N_m}{N_0r^{m-1}}.    
\end{equation}
It can be shown that $\hat{\tau}_{mlss}$ is an unbiased estimator of $\tau$. 
Intuitively, s-MLSS generates a forest of $N_0$ $r$-ary trees of sample paths with depth $m$:
the root is the initial state, nodes are the states at which we split, and edges are simulated sample paths.
The total number of leaf nodes is at most $N_0r^{m-1}$, and we count the total number of them, $N_m$, reaching the target.
In turn, it gives us the estimator in form $N_m / N_0r^{m-1}$.
See full proof in Appendix~\ref{appendix:proof}.

\begin{proposition}\label{lemma:mlss-unbias}
Under the ``no level-skipping'' assumption (\ref{eq:mlss-assumption}), using the MLSS with $m$ levels and a splitting ratio $r$, 
$\hat{\tau}_{mlss}$ is an unbiased estimator of $\tau$; that is, 
$\E[\hat{\tau}_{mlss}] = \tau$.
\end{proposition}

\mparagraph{Variance Analysis}
Assume that we have sampled and simulated $N_0$ independent root paths and denote the number of paths that hit the target level by $N_m$. The variance of our estimate is
\begin{equation}\label{eq:mlss-var}
    \Var(\hat{\tau}_{mlss}) = \Var(\frac{N_m}{N_0r^{m-1}}) = \frac{\Var(N_m)}{N_0^2 r^{2(m-1)}}.
\end{equation}
 Further denote the number of offsprings of the $k$-th root path that hit the target level by $N_m^{\langle k \rangle}$, $k=1, \ldots, N_0$. We have $\Var(N_m) = \sum_{i=1}^{N_0}\Var(N_m^{\langle i \rangle}) = N_0 \Var(N_m^{\langle 1 \rangle})$.
Combining together we have
\begin{equation}\label{eq:mlss-var-2}
    \Var(\hat{\tau}_{mlss}) = \frac{\Var(N_m^{\langle 1 \rangle})}{N_0 r^{2(m-1)}}.
\end{equation}
It is hard to derive an analytical expression for $\Var(N_m^{\langle 1 \rangle})$ in the multi-level splitting setting because there are many dependencies caused by splitting and sharing.
However, we can easily estimate $\Var(N_m^{\langle 1 \rangle})$ by MLSS's root path simulations themselves using the standard variance estimator
\begin{equation}\label{eq:mlss-std}
\sigma^2 = \frac{\sum_{i=1}^{N_0}(N_m^{\langle i \rangle} - \overline{N}_m)^2}{N_0-1},
\end{equation}
where $\bar{N}_m$ is the sample mean of target hits from simulations of $N_0$ root paths.
Combining it together, we have an unbiased estimation of the variance of MLSS estimator as follows:
$\widehat{\Var}(\hat{\tau}_{mlss}) = \frac{\sigma^2}{N_0r^{2(m-1)}}$.

\mparagraph{Parallel Computations}
Since the simulations of root paths are independent, it is straightforward to parallelize MLSS on a group of machines to further improve computation efficiency.
We monitor the progress of simulations and synchronize counters on the machines periodically to produce a running estimate; the procedure stops until the estimate reaches the desired accuracy level.

\mparagraph{Relationship between SRS and MLSS}\label{sec:mlss-vs-srs}
It is not hard to prove that SRS is a special case of MLSS with splitting ratio $r=1$.
As $r=1$, $\displaystyle\hat{\tau}_{mlss} = N_m/N_0 = \hat{\tau}_{srs}$. 
Similarly, 
$\Var(\hat{\tau}_{mlss}) = \Var(N_m^{\langle 1 \rangle})/N_0 = \tau_{srs}(1-\tau_{srs})/N_0$, that degenerates to $\Var(\hat\tau_{srs})$.

However, we still need careful considerations to apply MLSS in practice for the best performance; e.g., how to select splitting ratio $r$, how many partitions of levels  do we need and how to decide the boundaries of partitions.
There are many trade-offs among those choices.
We discuss how to solve for the optimal setting of MLSS that minimizes the simulation cost in Section~\ref{sec:optimal}.
\section{General MLSS}\label{sec:extension}
\ansb{
The last section introduced the s-MLSS estimator
under the ``no level-skipping'' assumption.
In general, this assumption can be easily violated, e.g., with volatile stochastic processes such as stock prices.
To remove this assumption and make MLSS more widely applicable, we propose a novel and general MLSS procedure.
}
Without the ``no level-skipping'' assumption, (\ref{eq:mlss-assumption}) and (\ref{eq:mlss-truth}) no longer hold and need to be modified. With the same sequence of boundaries $0 = \beta_0 < \beta_1 < \cdots < \beta_m = 1$, we denote by $U_i(SP) = \inf\{t \ge 0 \mid f(x_t) \ge \beta_i\}$ the first time that a sample path $SP: \{x_t\}_{t\ge 1}$ \emph{crosses} boundary $\beta_i$.
Notice the difference here between $U_i$ and $T_i$ (in Section~\ref{sec:mlp}) that $T_i$ specifically requires the process to land inside $L_i$ while $U_i$ only requires the process to pass the lower boundary of $L_i$.
Denote by $\Theta_i=\{SP \mid U_i(SP) \le s\}$ the event that the process crossed boundary $\beta_i$ before $s$. 
Similarly, we have 
\begin{equation}\label{eq:g-mlss-theta}
    \Theta_m \subset \Theta_{m-1} \subset \cdots \subset \Theta_1 \subset \Theta_0,
\end{equation}
and subsequently
\begin{align}\label{eq:g-mlss-prob}
\resizebox{.9\hsize}{!}{
    $\tau = \Pr{\Theta_m}
    = \Pr{\Theta_m \mid \Theta_{m-1}}\cdots\Pr{\Theta_1\mid \Theta_0}\Pr{\Theta_0} = \prod_{i=1}^m \pi_i 
    $
    }
\end{align}
where $\pi_i = \Pr{\Theta_i \mid \Theta_{i-1}}$. 
The above probability decomposition is general and carries no assumption.
Next we outline the g-MLSS sampling procedure that unbiasedly estimates $\pi_i$.

\subsection{g-MLSS Sampler and Estimator}\label{sec:mlss-general}
\begin{figure}
    \centering
    \vspace{-1em}
    \subfloat[normal path]{\includegraphics[width=0.24\textwidth]{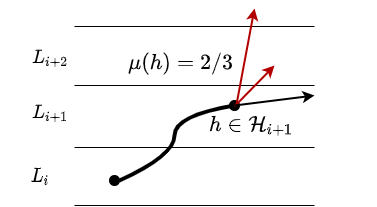}}
    \subfloat[level-skipping path]{\includegraphics[width=0.24\textwidth]{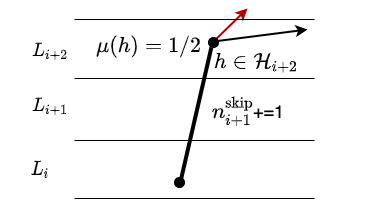}}
    \caption{Different types of simulated partial paths.}
    \label{fig:g-mlss}
\end{figure}
The g-MLSS sampler starts simulations from root paths, and recursively split the sample path whenever it \emph{lands} in \emph{any} level for the first time until time runs out or the path satisfies the query condition.
Whenever a splitting happens, we record the proportion of offspring processes that cross the higher boundary of the level in which the splitting happens.


Formally, in a realized g-MLSS simulation, denote by $\mathcal{H}_i \subset \States \times \Times$ the set of splitting states in $L_i$, each of which belongs to a separate path that lands in $L_i$ before $s$. 
On the other hand, denote by $n_i^{\text{skip}}$ the number of paths that pass $\beta_{i+1}$ \emph{without} landing in $L_i$, where level-skipping happens (see Figure \ref{fig:g-mlss}). For any $h\in \mathcal{H}_i$, let $\mu(h)$ denote the ratio of $h$'s offsprings that cross $\beta_{i+1}$. 
The estimator of $\pi_{i+1}, i > 0$ is given by
\begin{equation}\label{eq:level-cross-prob}
    \hat{\pi}_{i+1}=\frac{1}{|\mathcal{H}_i| + n_i^{\text{skip}}}\Big(\sum_{h\in \mathcal{H}_i}\mu(h) + n_i^{\text{skip}}\Big).
\end{equation}
\ansa{%
It is worth noting that, with g-MLSS, there is no need to use a unified splitting ratio $r$ as in s-MLSS.
After all, when a path splits at $h$, we only need the ratio $\mu(h)$ instead of the size of its offsprings.
}

The special case is the starting level $L_0$, since we directly start with $N_0$ independent root paths.
The estimation of $\pi_1$ is given by $\displaystyle \hat{\pi}_1 = \frac{\card{\mathcal{H}_1} + n_1^{\text{skip}}}{N_0}$.
Following the probability decomposition by (\ref{eq:g-mlss-prob}), the general MLSS estimator for $\tau$ is
\begin{equation}\label{eq:general-mlss-estimate}
    \hat{\tau}_{mlss} = \prod_{i=1}^m \hat{\pi}_i.
\end{equation}

\begin{proposition}\label{lemma:general-mlss}
In general, using the Multi-Level Splitting Sampling with $m$ levels, $\hat{\tau}_{mlss}$ in (\ref{eq:general-mlss-estimate}) is an unbiased estimator of $\tau$; that is, $\E[\hat{\tau}_{mlss}] = \tau$.
\end{proposition}

Given the general form of MLSS as above, it is more clear how s-MLSS is a special case of the general one.
With the ``no level-skipping'' assumption, $n_i^{\text{skip}}$'s are always zero.
Given a unified splitting ratio $r$, for any splitting state $h \in \mathcal{H}_i$ in $L_i$, 
$\mu(h) = N_{i+1}(h)/r$, where $N_{i+1}(h)$ denotes the number of hits (from $h$'s split offsprings) to hit the next level $L_{i+1}$.
Additionally, $\card{\mathcal{H}_i} = N_{i}$ (recall that $N_i$ is the number of entrances to $L_i$).
Hence, (\ref{eq:level-cross-prob}) degenerates to
$\displaystyle \hat{\pi}_{i+1} = \frac{1}{N_i}\frac{\sum_{h \in \mathcal{H}_i}N_{i+1}(h)}{r} = \frac{N_{i+1}}{rN_i}$,
and the g-MLSS estimator by (\ref{eq:general-mlss-estimate}) is equivalent to the s-MLSS estimator by (\ref{eq:mlss-est}).

In summary,  g-MLSS greatly extends the applicability of s-MLSS by allowing level-skipping and a dynamic splitting ratio. 
With the g-MLSS algorithm and estimator, we are able to efficiently obtain quality estimations of durability prediction queries on mostly any temporal process that exhibits continuity and temporal dependence.
\ansa{%
Moreover, the flexible splitting procedure opens up many interesting opportunities for optimization, e.g., how to optimally allocate splitting ratios across sample paths, or how to learn and converge to the optimal assignment on-the-fly while conducting MLSS.
}
\begin{figure}
    \centering
    \includegraphics[width=0.2\textwidth]{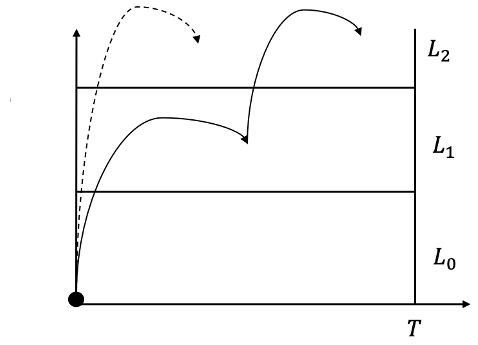}
    \caption{A simple two-level case with level-skipping. \mdseries Dashed path represents a (discrete-time) series that directly goes from $L_0$ to $L_2$ skipping $L_1$.}
    \label{fig:level-skipping}
\end{figure}
\subsection{Variance Analysis}\label{sec:g-mlss-var}
In order to practically apply g-MLSS, we need its variance term to determine the stopping condition (i.e., confidence interval or relative error) for durability query processing.
The variance of general MLSS estimator as in (\ref{eq:general-mlss-estimate}) would be very complicated and challenging, because the underlying stochastic process takes general forms.
Unfortunately, we do not have a closed-form expression of the variance for the general case in this paper.
With that being said, it will not limit the utility of g-MLSS in practice. 
\ansa{
In this section, we first showcase the variance analysis of general MLSS estimator for a simple but non-trivial case---two levels with level-skipping (Figure~\ref{fig:level-skipping}).
Then, we show how to use \emph{bootstrapping}~\cite{efron1994introduction} to provide a variance estimate of g-MLSS for the general case in practice.
}
\mparagraph{Simple two-level Level-skipping}\label{sec:g-mlss-2-level}
\ansa{
There are two types of paths hitting the target: solid line path (with no level-skipping) and dashed line path (directly jump from $L_0$ to $L_2$).
With abuse of notation, let $p_{0,1} = p_1$ and $p_{1,2} = p_2$ (recall (\ref{eq:mlss-assumption}) and (\ref{eq:mlss-truth})).
The numbers in subscript simply represent the transition between levels.
These probabilities represent the normal case as we discussed in Section~\ref{sec:mlss}. 
However, with the existence of level-skipping, we need to introduce an additional probability $p_{0,2}$ denoting the chance of level-skipping. 
Hence, the ground truth hitting probability consists of two parts:
$\tau = p_{0,1} p_{1,2} + p_{0,2}$.
Accordingly, we decompose counter $N_2$ (number of hits to the target) as $\smash{N_2 = N_2^{(ns)} + N_2^{(s)}}$, where $N_2^{(ns)}$ is the number of hits from non-skipping paths while $\smash{N_2^{(s)}}$ is the number of hits from level-skipping paths.
Then, our estimator also consists of the estimations of these two parts,
$\hat{\tau}_{mlss} = N_2^{(ns)}/N_0  r + N_2^{(s)}/N_0$.
For variance, we have
\(Var(\hat{\tau}_{mlss}) = \Var\big(N_2^{(ns)}\big)/N_0^2r^2 + \Var\big(N_2^{(s)}\big)/N_0^2\).
First, $N_2^{(s)}$ can be viewed as a binomial variable with $N_0$ trials and probability $p_{0,2}$; i.e., $\smash{N_2^{(s)} \sim B(N_0, p_{0,2})}$.
Thus $\smash{\Var\big(N_2^{(s)}\big) = N_0p_{0,2}(1 - p_{0,2})}$.
Second, the quantity $\smash{N_2^{(ns)}}$ conditions on the number of paths without skipping ($N_1$), which it is also an random variable throughout the sampling procedure.
We cannot just break it up as in standard variance analysis. 
Instead, we should do a conditioning on number of non-skipping paths and use the law of total variance:

\begin{small}
\begin{align*}
    \begin{split}
        \Var\big(N_2^{(ns)}\big) &= \Var\bigg(\E[N_2^{(ns)} \mid N_1]\bigg) + \E\bigg[\Var(N_2^{(ns)} \mid N_1)\bigg] \\
        &= \Var(N_1rp_{1,2}) + \E[N_1\Var(N_2^{\langle 1 \rangle})]\\
        &~~~\text{($\Var(N_2^{\langle i \rangle}) = \Var(N_2^{\langle j \rangle}) = \Var(N_2^{\langle 1 \rangle})$, for $\forall i,j<N_0$)}\\ 
        &= r^2p^2_{1,2}\Var(N_1) + \E[N_1]\Var(N_2^{\langle 1 \rangle}) \\
        &= r^2p^2_{1,2}N_0p_{0,1}(1 - p_{0,1}) + \E[N_1]\Var(N_2^{\langle 1 \rangle}).\\ &~~~\text{($N_1 \sim B(N_0, p_{0,1})$)}
    \end{split}
\end{align*}
\end{small}
\noindent Hence, 
\[
\resizebox{.8\hsize}{!}{
    $\frac{\Var\big(N_2^{(ns)}\big)}{N_0^2r^2} = p^2_{1,2}\frac{p_{0,1}(1 - p_{0,1})}{N_0} + p_{0,1}\frac{\Var(N_2^{\langle 1 \rangle})}{N_0r^2}$
}.
\]
Putting it all together, we have
\begin{equation}\label{eq:mlss-general-var}
\begin{split}
\resizebox{.9\hsize}{!}{
    $\Var(\hat{\tau}_{mlss})
    = p^2_{1,2}\frac{p_{0,1}(1 - p_{0,1})}{N_0} + p_{0,1}\frac{\Var(N_2^{\langle 1 \rangle})}{N_0r^2} + \frac{p_{0,2}(1 - p_{0,2})}{N_0}$
    }.
\end{split}
\end{equation}
In practice, we can use $\hat{p}_{0,1} = N_1/N_0$ as an unbiased estimation for $p_{0,1}$. 
Similarly, $\smash{\hat{p}_{0,2} = N_2^{(s)}/N_0}$ and $\smash{\hat{p}_{1,2} = N_2^{(ns)}/N_1r}$ as unbiased estimations for $p_{0,2}$ and $p_{1,2}$, respectively.
$\smash{\Var(N_2^{\langle 1 \rangle})}$ can be estimated unbiasedly similar to (\ref{eq:mlss-var-2}) by reusing the simulated root paths.
Again, it is not hard to find that the variance term we derived in (\ref{eq:mlss-var-2}) is a special case of the above equation when there is no level-skipping; i.e., $p_{0,2} = 0$ and $p_{0,1} = 1$.

We believe the variance analysis of g-MLSS estimator, though very complex, would follow the similar procedure as in the simple two-level case.
We leave this part as one of the future work.
}

\mparagraph{General Level-skipping and Bootstrapping Evaluation}\label{sec:g-mlss-bootstrap}
\ansa{
In the general case, the standard technique of bootstrap sampling can be used in practice to provide a good estimation for variance of g-MLSS estimator.
This approach is widely used to empirically estimate the variance or the distribution of sample mean when the population variance is complex and or not accessible.

More specifically, in our setting, assume we have already simulated $N_0$ root paths and obtained an estimate $\hat{\tau}_0$ of the hitting probability. 
In one bootstrap run, we randomly draw $n$ root paths, \emph{with replacement}, from the existing root paths, from which we calculate a g-MLSS estimate, called a bootstrap estimate.  We perform $N$ such independent bootstrap runs to obtain $N$ bootstrap estimates $\hat{\tau}_i,i=1,...,N$. 
From the empirical distribution of $\hat{\tau}_i,i=1,...,N$, we calculate the (bootstrapped) variance for g-MLSS, i.e.,
$\widehat{Var}(\hat{\tau}_0) = \sum_{i=1}^{N}(\hat{\tau}_i - \bar{\tau})^2 / N$, where $\bar{\tau}$ is the mean of bootstrap estimates from $N$ bootstrap runs.

Despite the simplicity and effectiveness of bootstrap sampling, it may incur considerable evaluation cost, as bootstrapping essentially replays the history multiple times, and may become increasingly expensive as we expand the sample pool.
Compared to the case of s-MLSS or g-MLSS with the two-level setting where variance can be directly calculated, applying g-MLSS with bootstrap evaluation requires more care to achieve good overall performance.
There are several techniques for speeding up bootstrap sampling, ranging from more advanced subsampling procedures~\cite{kleiner2014scalable,basiri2015robust} to parallel computation.
A practical rule of thumb that we found in practice is to run bootstrap evaluation conservatively---compared with frequent bootstrapping to ensure that we never overshoot the given quality target, sometimes overrunning the simulation a little would be overall more efficient.
As we will see in Section~\ref{sec:expr}, applying this rule, even with an unoptimized bootstraping implementation, g-MLSS can still provide up to 5x overall speedup over SRS.
}

\section{Optimizing MLSS Design}\label{sec:optimal}

There is still one missing piece in applying MLSS: how do we choose its parameters?
Specifically, how many levels do we need, and given the number of levels, how do we properly partition the value function range into levels?
In Section~\ref{sec:extension}, we saw that g-MLSS also allows variable splitting ratios---how do we additionally choose these?
Manually tuning all these parameters is clearly impractical.
On the other hand, automatic optimization is also challenging because of the vast space of possibilities as well as the difficulty of not knowing the effectiveness of our choices a priori.

We make some simplifying assumptions to make the optimization problem tractable.
1)~We focus only on choosing level partition plans; we forgo the freedom of setting variable splitting ratios and instead choose a small, fixed splitting ratio $r$.
As validated in experiments in Section~\ref{sec:expr}, large ratios tend to be suboptimal because they dramatically increase the number of paths at higher levels.
Furthermore, variable splitting ratios can be effectively approximated by replacing a level having a large ratio with multiple levels, each having the same small fixed ratio.
%
\ansa{
2)~We derive an empirical measure for evaluating different MLSS parameter settings (Section~\ref{sec:plan-eval}).
For ease of derivation and measurement, we make the same no level-skipping assumption as in s-MLSS.
While this assumption does not hold in general, it allows to obtain a surrogate measure that is much cheaper to estimate.
Importantly, it does not affect the correctness of our sampling and estimation procedures in any way because it is only used to choose a partition plan.
Moreover, as we will see in experiments in Section~\ref{sec:expr:mlss-opt}, this measure work well in practice for both s-MLSS and g-MLSS on various models and query types.}
%
3)~We then present an adaptive greedy strategy (Section~\ref{sec:greedy}) that searches for the parameters settings aimed at optimizing the above empirical metric.
\ansa{
This strategy is generic, applicable to both s-MLSS and g-MLSS, and can work with other, better empirical measures if available.
}

\subsection{Partition Plan Evaluation}\label{sec:plan-eval}
Previous work in statistics~\cite{l2006splitting} showed an analogy between MLSS (with fixed splitting ratio) and branching process theory~\cite{harris1964theory}, and concluded that the optimal setting for MLSS is to make advancement probabilities between consecutive levels roughly the same, called a ``balanced growth.''
That is, consider MLSS with $m$ levels, 
\begin{equation}\label{eq:balance-growth}
    p_1 = p_2 = \cdots = p_m = p = \tau^{1/m}.
\end{equation}
From standard branching process theory, we have
\begin{equation}\label{eq:branch-var}
    \Var(\hat{\tau}_{mlss}) = \frac{m(1-p)p^{2m-1}}{N_0}.
\end{equation}
The above expression indicates that given a fixed number of root paths, more levels lead to smaller variance.
However, more levels also lead to more expensive simulation cost of a root path because of the exponential splitting growth of a root path through the levels.
Our optimization goal, using MLSS as approximate query processing technique, is not just to minimize variance.
Instead, we hope to minimize the variance in a fixed amount of the time.
Ultimately, the query time using MLSS is determined by the variance of the estimator.
A smaller variance in unit time directly leads to less simulation cost for answering durability query.
Though the ``balanced growth'' strategy is a reasonable guideline to partition the space, it is still not clear, in practice, how to partition levels that create balanced growth and how to choose the right number of levels.

To meet our needs, we propose the following evaluation metric.
Consider a level partition plan $B$, which consists of a set of values what we call ``partition boundaries''; that is, 
$B = \{v  \mid v \in (0, 1)\}$.
Given a fixed amount of simulation budget, say $t_0$ time, we have simulated $N(t_0)$ root paths (including all its splitting copies).
Note that $N(t_0)$ is a random variable depending on the total time $t_0$ and the average simulation time $c_B$ of a root path using partition plan $B$.
We define an evaluation function for $B$ in terms of variance of estimator $\hat{\tau}_{mlss}$ by $N(t_0)$:
\begin{equation}\label{eq:eval}
    eval(B) = \Var\bigg(\frac{N_m(t_0)}{N(t_0) r^{m-1}}\bigg),
\end{equation}
where $N_m(t_0)$ is a random variable denoting the total number of target hits within $t_0$ time.
In this case, $m$ = \card{B} + 1, denoting the total number levels induced by plan $B$.
Since $N(t_0)$ is a random variable, we should express the variance term conditioning on $N(t_0)$, and use the law of total
variance and the decomposition trick in (\ref{eq:mlss-var-2}): $eval(B) = \E\bigg[\frac{\Var(N_m^{\langle 1 \rangle})}{N(t_0) r^{2(m-1)}} \mid N(t_0)\bigg] + \Var\bigg(\E\bigg[\frac{N_m(t_0)}{N(t_0) r^{m-1}} \mid N(t_0)\bigg]\bigg)$.

Given $N(t_0)$, $\displaystyle \E\big[\tfrac{N_m(t_0)}{N(t_0) r^{m-1}}\big] = \tau$, thus the second term in the above equation is 0.
Recall that $N_m^{\langle 1 \rangle}$ is a random variable denoting the number of target hits from a root path.
Given $N(t_0)$ and partition plan $B$, $\Var(N_m^{\langle 1 \rangle})$ becomes a constant.
Hence, the first term in the equation becomes
$\displaystyle \tfrac{\Var(N_m^{\langle 1 \rangle})}{r^{2(m-1)}}\E[1/N(t_0)]$.
The term $\E[1/N(t_0)]$ can be roughly estimated by $\displaystyle 1 / \tfrac{t_0}{c_B} = c_B/t_0$.
Finally, the evaluation function of a partition plan $B$ is
\begin{equation}\label{eq:eval-final}
    eval(B) = \frac{\Var(N_m^{\langle 1 \rangle})}{r^{2(m-1)}}\frac{c_B}{t_0}.    
\end{equation}

Ideally, given a fixed amount of time $t_0$, we hope to solve for partition plan $B$ that minimizes the objective $eval(B)$.
However, this optimization problem is hard to solve analytically, since $\Var(N_m^{\langle 1 \rangle})$ and $c_B$ are themselves variables when the plan $B$ changes.
But fortunately, we can optimize the objective empirically, as $\Var(N_m^{\langle 1 \rangle})$ and $c_B$ can be estimated through the MLSS simulations.
$\Var(N_m^{\langle 1 \rangle})$ can be estimated using variances of target hits from all simulated root paths,
and $c_B$ can also be estimated simply dividing $t_0$ by the number of simulations of root path within time $t_0$.
In this way, we can start with a candidate pool of partition plans.
For each candidate, we run MLSS simulations for the same amount of time $t_0$ as trial runs to estimate the objective $eval(B)$, and finally pick the best candidate that produces the minimum value.

\subsection{An Adaptive Greedy Partition Strategy}\label{sec:greedy}
To empirically optimize MLSS, it is prohibitively expensive to run trial simulations for all feasible partition plans and splitting ratios.
In this section, we present a heuristic greedy strategy that works well in practice to automatically search for (near-) optimal MLSS parameters.

\begin{algorithm}[t]\small
\newcommand\mycommfont[1]{\rmfamily{#1}}
\SetCommentSty{mycommfont}
\SetKwInOut{Input}{Input}
\SetKwInOut{Output}{Output}
\Input{Interval $I = [0,1]$.}
\Output{A partition plan $B = \{v \mid v \in (0, 1)\}$.}
\BlankLine
$B \leftarrow \emptyset$\;
$opt\_eval \leftarrow \mathtt{INT\_MAX}$\tcp*{remember the minimum so far}
$v_{lo} \leftarrow 0, v_{hi} \leftarrow 1$\;
\For{round $i = \{1,2,\cdots\}$} {
   Uniformly generate a value set $C = \{v \mid v \in (v_{lo}, v_{hi})\}$ as candidates for the $i$-th partition boundary\;
   $\displaystyle e^* = \min_{v \in C} eval(B \cup v)$\;
   $\displaystyle v^* = \argmin_{v \in C} eval(B \cup v)$\; 
   \uIf {$e^* < opt\_eval$} {
        $B \leftarrow B \cup v^*$\;
        $opt\_eval \leftarrow e^*$\;
        Find the level $[\beta_i, \beta_j] (\beta_i, \beta_j \in B, \beta_i < \beta_j)$, induced by $B$ and $I$, that has the smallest level advancement probability $p_{i,j}$\;
        $v_{lo} \leftarrow \beta_i, v_{hi} \leftarrow \beta_j$\;
   }\Else{
        break;
   }
}
\Return $B$\;
\caption{\label{algo:greedy}Adaptive Greedy Partition.}
\end{algorithm}
The main idea of our strategy is to adaptively and recursively partition the space---place the partition boundaries one by one and always partition the level with smaller level advancement probability.
The intuition behind our greedy behavior is two-fold: (1) A level with smaller level advancement probability means that this level is an ``obstacle'' blocking sample paths reaching the target. 
Partition such levels would focus the simulation resources more on success paths;
(2) as we recursively bisect levels with smaller advancement probability, it automatically moves towards a ``balanced growth'' situation where advancement probabilities from all levels are roughly the same.
Recall our discussion in Section~\ref{sec:plan-eval}; this behavior has already been confirmed by~\cite{l2006splitting} to have a better sampling efficiency.

Full description is shown in Algorithm~\ref{algo:greedy}.
Throughout the procedure, we adaptively make two decisions: the optimal number of levels, and the placement of these levels.
At the beginning, we start with the original interval $[0, 1]$ (Line 1).
Then we place the partition boundary one by one, recursively bisecting the value intervals, until a stopping condition is met (Line 4-14).
In the loop, we first generate a set of candidate boundaries (Line 5) and then use the empirical evaluation approach, as elaborated in Section~\ref{sec:plan-eval}, to find the optimal partition boundary (Line 6 and Line 7).
Finally, we need to update our partition plan and decide when to stop the procedure.
If the current best evaluation is better the previous, we continue to add a new partition boundary (Line 8-12).
Note that here we need to greedily pick the next interval with smallest advancement probability to partition (Line 11-12).
Otherwise, if the current best evaluation is already worse than the previous, there is no need to further add more partition boundaries to the plan, since more levels lead to exponential growth of splitting paths and would incur more expensive simulation cost overall.

The aforementioned empirical optimization framework saves users from the time-consuming manual parameters tuning process when applying MLSS in practice.
We set a reasonable fixed splitting ratio (which we further justify in Section~\ref{sec:expr:mlss-opt}) in advance, and our optimization framework will take care of the rest.
An additional benefit of our empirical optimization solution is that all trial runs of MLSS are not ``wasted.''
Since each trial simulation, no matter which plan it follows, returns an unbiased estimator.
So in process of picking the optimal parameters, we also are building up towards a reliable estimation for the query.

\section{Experiments}\label{sec:expr}
\begin{figure}
    \centering
    \includegraphics[width=0.3\textwidth]{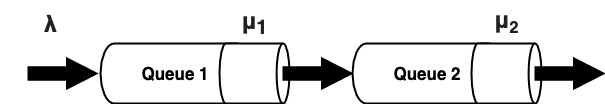}
    \vspace{1ex}
    \caption{Tandem Queue with Poisson arrivals and Exponential service time.}
    \label{fig:tandem-queue}
\end{figure}
We select three stochastic temporal processes with simulation models that are commonly used in practical applications.
 
\noindent\textbf{(1) Tandem Queues}: As shown in Figure~\ref{fig:tandem-queue}, we have a queueing system with tandem queues, which is the simplest non-trivial network of queues in \emph{queueing theory}~\cite{cooper1981queueing}.
The process is the following.
Customers come into Queue 1 following a Poisson distribution with $Pois(\lambda)$.
Queue 1 services each customer following an Exponential distribution with $Exp(\mu_1)$, and then sends customers into Queue 2.
Queue 2 services each customer by another Exponential distribution with $Exp(\mu_2)$ before customers leave the system.
We consider the number of customers in Queue 2 as a stochastic process, and always start with an empty system (i.e., two empty queues).
In our experiments, we set $\lambda=0.5, \mu_1 = \mu_2 = 2$.
Though in simple form, queuing system is the foundation for many real-world problems~\cite{newell2013applications}, e.g., birth-death process, supply chains, transportation scheduling, and computer networks analysis~\cite{lazowska1984quantitative}.
Durability queries on such models are widely used to evaluate the robustness of systems.

\noindent\textbf{(2) Compound-Poisson Process:}
A Compound-Poisson Process (CPP) can be described by the following stochastic process $U = \{U(t)\}_{t\ge 0}$.
$U(t) = u + ct - S(t)$, where $S(t)$ is a compound Poisson process with jump density $\lambda$ and jump distribution $F$, and $u, c>0$ are constants.
This type of process is commonly used in the financial world for risk management and financial product design~\cite{albrecher2008levy}.
Intuitively, imagine a insurance policy with $u$ as initial surplus and $c$ as users' monthly payment.
The compound Poisson process $S(t)$ represents the aggregate claim payments up to time $t$.
Then the overall stochastic process $U$ shows the net profit of this insurance policy.
In our experiments, we set Poisson jump density $\lambda=0.8$ and use uniform distribution $Uni(5, 10)$ as jump distribution.
We choose $u=15$ and $c=4.5$.
Recall durability query examples in Section~\ref{sec:intro}, durability queries in financial domains can be used for revenue projection, risk management, and financial product design.

\noindent\textbf{(3) Recurrent Neural Networks:}
\begin{figure}
    \centering
    \includegraphics[width=0.3\textwidth]{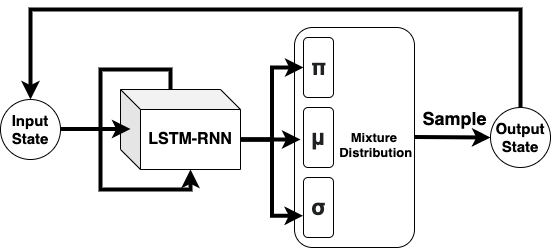}
    \caption{A stochastic process by LSTM-RNN-MDN.}
    \label{fig:rnn}
\end{figure}
As shown in Figure~\ref{fig:rnn}, we train a Recurrent Neural Networks (RNN) with Long-Short Term Memory (LSTM) and Mixture Density Network (MDN)~\cite{bishop1994mixture} using Google's 5-year daily stock prices from 2015 to 2020.
The LSTM-RNN-MDN structure has proven its success at many real-life tasks of probabilistically modeling and generating sequence data: e.g., language models~\cite{bengio2003neural}, speech recognition~\cite{graves2013speech}, hand-writing analysis~\cite{graves2009offline}, and music composition~\cite{10.5555/870511}.
In our network, we use two stacked RNN layers, 256 LSTM units per RNN layer, and a 2-dimensional mixture layer with 5 mixtures.
During training phase, we trained the model for sequence length of 50, and for 100 epochs with a batch size of 32.
Such learning-based black-box model demonstrates the general applicability of MLSS and of durability queries.

\mparagraph{Evaluation Metric}
We evaluate the performance of different methods using the following two metrics: total number of simulation steps (invocations of simulation procedure $\SimStep$) and total simulation time.
In our experiments, we run sampling procedures until the estimation satisfies a given quality target.
Specifically, we use two quality measurements throughout our experiments:

\noindent\textbf{(1) Confidence Interval:} Confidence interval (CI) is a statistical measurement for point estimates.
    It shows how likely (or how confident) that the true parameter is in the proposed range.
    There is no universal formula to
    construct CI for an arbitrary estimator.
    However, if a point estimator $\hat{\mu}$ takes the form of the mean of $n$ independent and identically distributed (i.i.d.) random variables with equal expectation $\mu$, then by the Central Limit Theorem and Normal Approximation, an approximate 1-$\alpha$ CI of $\mu$ can be constructed by:
   $\displaystyle
    \smash{[\hat{\mu} - z_{\alpha/2}\sqrt{\sigma^2/n}, \hat{\mu} + z_{\alpha/2}\sqrt{\sigma^2/n}]},
    $
    where $z_{\alpha/2}$ is the Normal critical value with right-tail probability $\alpha/2$, and $\sigma^2$ is the variance of estimate.
    By default, to obtain reliable query answers, we require that all estimations should have a 1\% CI with 95\% confidence level (i.e., $z_{\alpha/2} = 1.96$).
    Unfortunately, the standard CI, as in the above equation, has a limitation: when the true probability $\mu$ is very close to 0 or 1, where the Normal Approximation assumption does not hold, the CI guarantee would break.
    Hence, we also consider another quality measurement below for extreme cases.

\noindent\textbf{(2) Relative Error:}
    Relative Error (RE) measures the variance (of estimate) as a relative ratio to the true probability, defined as follows:
    $\displaystyle
        \smash{RE = \sqrt{\sigma^2}/\mu},
    $
    where $\mu$ is the true probability and $\sigma^2$ is the variance of estimate.
    This is not feasible to calculate directly in practice, since we do not know the true probability $\mu$ before the query.
    But in practice, we can roughly estimate the ground truth probability, and use that to fairly compare the RE ratio among different methods.
    By default, we require that all estimations should have a low relative error at 10\%.
    Unlike CI, RE is widely applicable to any scenario.

In sum, throughout the experiment section, we evaluate durability queries with different ground truth probabilities.
For queries that have small-to-moderate probability (i.e., $>0.05$), we use Confidence Interval as the quality measure.
For queries that have tiny probability (i.e., $10^{-4}$ to $10^{-2}$), we use Relative Error as the alternative measure.
\mparagraph{Implementation Details}
All stochastic temporal models and proposed solutions were implemented in Python3.
More specifically, for neural network's construction and training, we use Keras~\cite{chollet2015keras} (back end by TensorFlow~\cite{tensorflow2015-whitepaper}).
\ansb{
Unless otherwise stated, for MLSS (s-MLSS and g-MLSS), to limit the number factors influencing performance, by default we fix the splitting ratio $r=3$ and use ``balanced-growth'' level-partition plans (recall discussions in Section~\ref{sec:plan-eval}), which are obtained by manual tuning given the number of partitions.
Effectiveness of the adaptive greedy partition strategy will be examined separately in Section~\ref{sec:expr:mlss-opt}.
For durability queries, we use the query condition in the form $z(x_t) \ge \beta$ and the simple value function $f(x_t) = \min\{z(x_t)/\beta, 1\}$ as we introduced in Section~\ref{sec:mlss}. 
Recall that $z(\cdot)$ is a real-valued evaluation of a state and $\beta$ is a user-specified value threshold.
For Queue model, $z(\cdot)$ evaluates the state by returning the number of customers in Queue~2; for CPP model, it is the value of $U(t)$, and for RNN model, $z(\cdot)$ returns the (simulated) stock price at a given state.
}
All experiments were performed on a Linux machine with two Intel Xeon E5-2640
v4 2.4GHz processor with 256GB of memory.

\begin{table}\small
    \centering
    \caption{Query settings on different models.}
    \label{tab:query-setting}
    \tabcolsep=0.11cm
    \begin{tabular}{|c|c|c|c|c|}
    \hline
        Query Type & Medium ($s, \beta$) & Small ($s, \beta$) & Tiny ($s, \beta$) & Rare ($s, \beta$)  \\\hline
        Queue Model & 500, 20 & 500, 26 & 500, 40 & 500, 45 \\\hline
        CPP Model & 500, 300 & 500, 350 & 500, 450 & 500, 500 \\\hline
        RNN Model & - & 200, 1550 & 200, 1600 & - \\\hline
    \end{tabular}
\end{table}
\begin{table}\small
    \centering
    \caption{Query answer comparisons on Queue Model. \mdseries
    Results are averaged over 100 runs with standard deviation.}
    \label{tab:ans-queue}
    \tabcolsep=0.1cm
    \begin{tabular}{|c|c|c|c|c|}
        \hline
        Query Type & Medium & Small & Tiny & Rare \\\hline
        SRS & \makecell{17.2\%$\pm$0.5\%} & \makecell{5.1\%$\pm$0.5\%} & \makecell{0.15\%$\pm$0.03\%} & \makecell{0.04\%$\pm2e^{-5}$} \\\hline
        MLSS & \makecell{17.9\%$\pm$0.4\%} & \makecell{5.5\%$\pm$0.5\%} & \makecell{0.17\%$\pm$0.02\%} & \makecell{0.04\%$\pm3e^{-5}$} \\\hline
    \end{tabular}
\end{table}
\begin{table}\small
    \centering
    \caption{Query answer comparisons on CPP Model. \mdseries
    Results are averaged over 100 runs with standard deviation.}
    \label{tab:ans-cpp}
    \tabcolsep=0.1cm
    \begin{tabular}{|c|c|c|c|c|}
        \hline
        Query Type & Medium & Small & Tiny & Rare \\\hline
        SRS & \makecell{15.5\%$\pm$0.5\%} & \makecell{5.3\%$\pm$0.5\%} & \makecell{0.24\% $\pm$0.02\%} & \makecell{0.03\%$\pm3e^{-5}$} \\\hline
        MLSS & \makecell{15.6\%$\pm$0.4\%} & \makecell{5.3\%$\pm$0.5\%} & \makecell{0.26\%$\pm$0.01\%} & \makecell{0.03\%$\pm4e^{-5}$} \\\hline
    \end{tabular}
\end{table}
\begin{table}\small
    \centering
    \caption{Query performance (single run) on RNN Model.}
    \label{tab:ans-nn}
    \tabcolsep=0.1cm
    \begin{tabular}{|c|c|c|}
        \hline
        Query Type & Small & Tiny \\\hline
        SRS & \makecell{2.6\%, 3.8 hours \\ 1,009,431 steps} & \makecell{0.51\%, 33.7 hours \\ 7,262,735 steps}\\\hline
        MLSS & \makecell{1.9\%, 0.75 hour \\ 196,913 steps}  & \makecell{0.45\%, 3.9 hours \\ 804,035 steps}\\\hline
    \end{tabular}
\end{table}
\begin{figure}[t]
    \centering
    \subfloat[Simulation steps]{\includegraphics[width=0.24\textwidth]{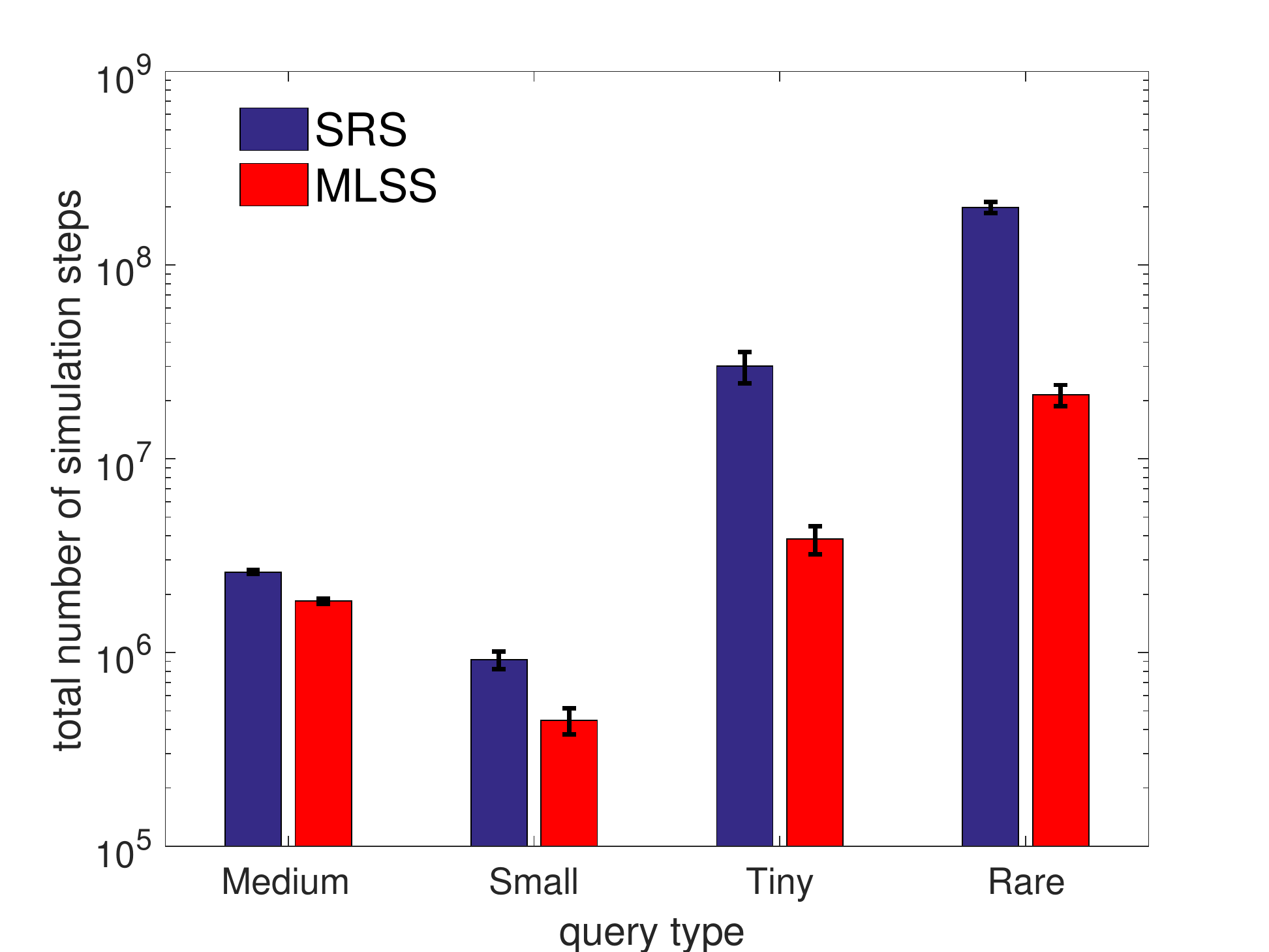}}
    \subfloat[Query time]{\includegraphics[width=0.24\textwidth]{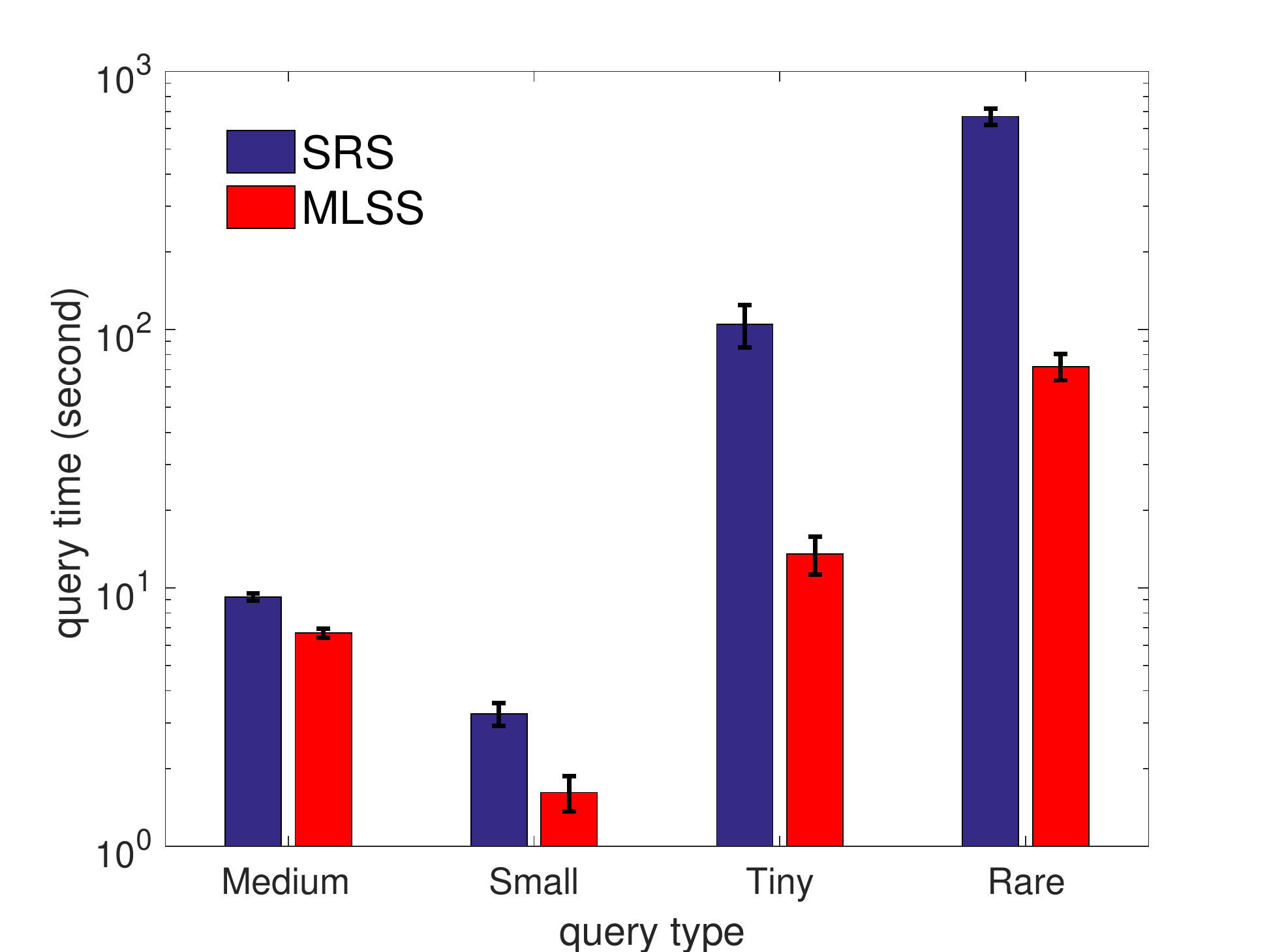}}
    \caption{Query efficiency on Queue Model}
    \label{fig:queue-efficiency}
\end{figure}
\begin{figure}[t]
    \centering
    \subfloat[Simulation steps]{\includegraphics[width=0.24\textwidth]{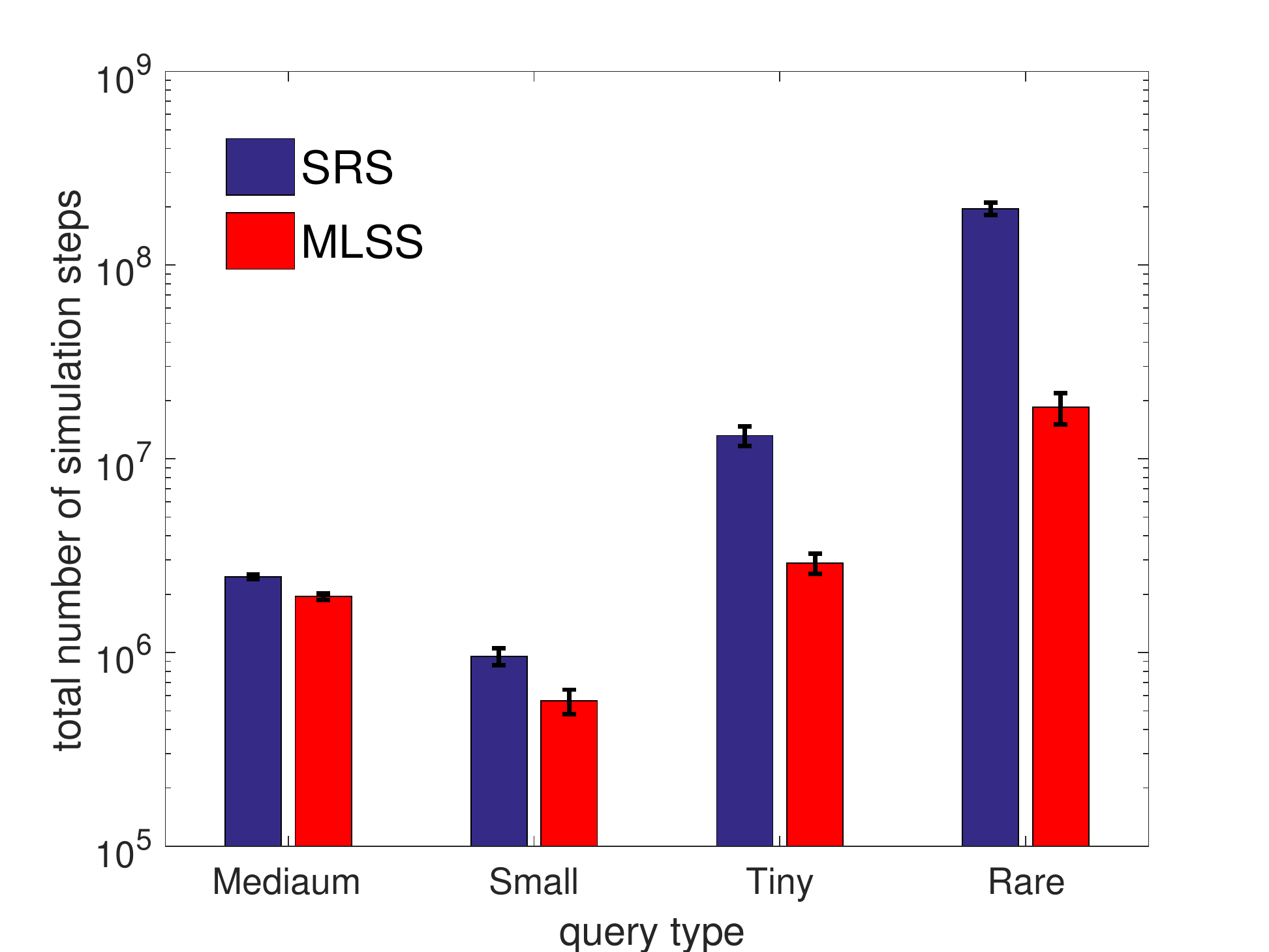}}
    \subfloat[Query time]{\includegraphics[width=0.24\textwidth]{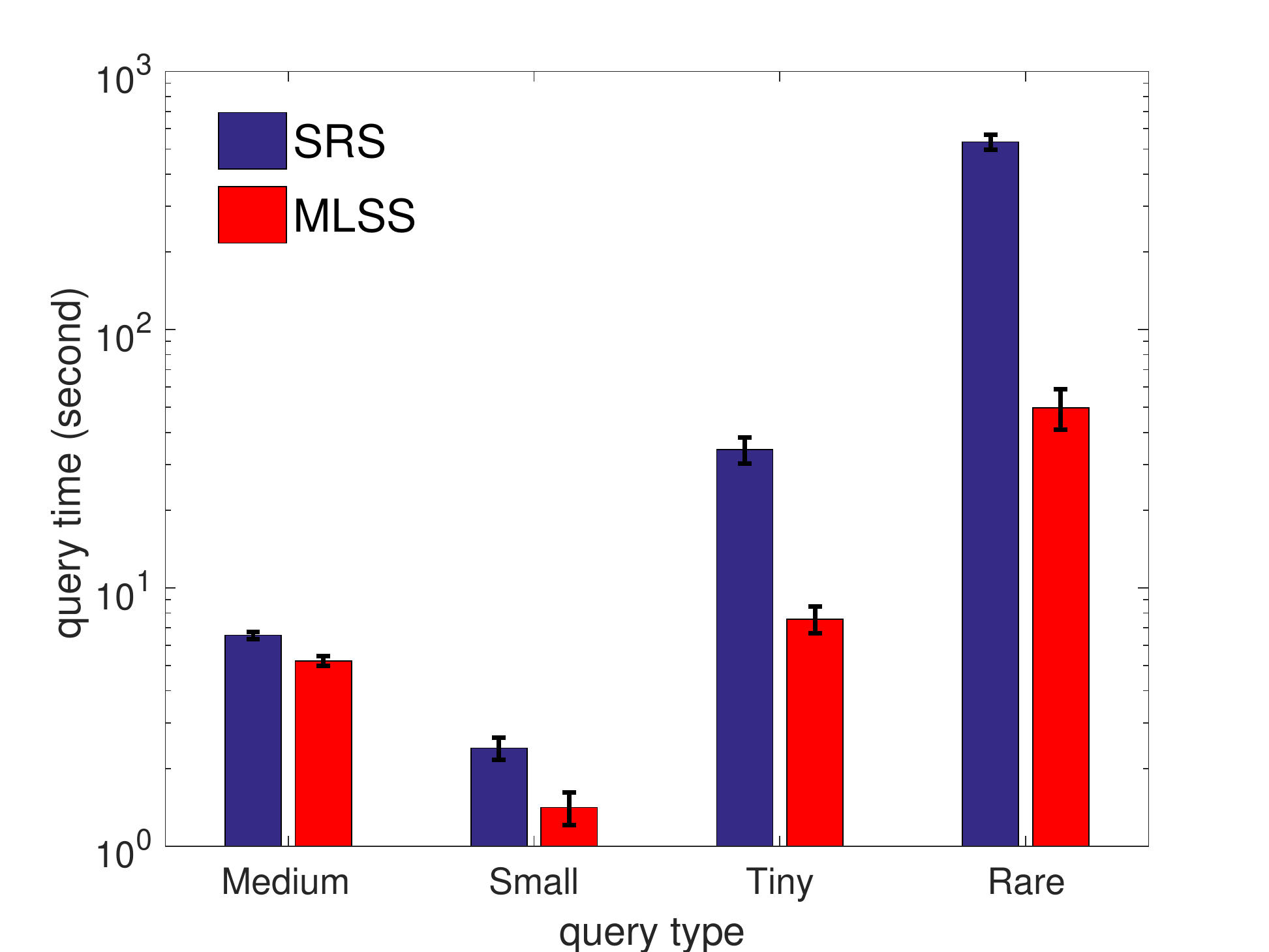}}
    \caption{Query efficiency on CPP Model}
    \label{fig:cpp-efficiency}
\end{figure}

\begin{figure*}
\centering
\subfloat[Queue Model, Small Query, CI]{\includegraphics[width=0.3\textwidth]{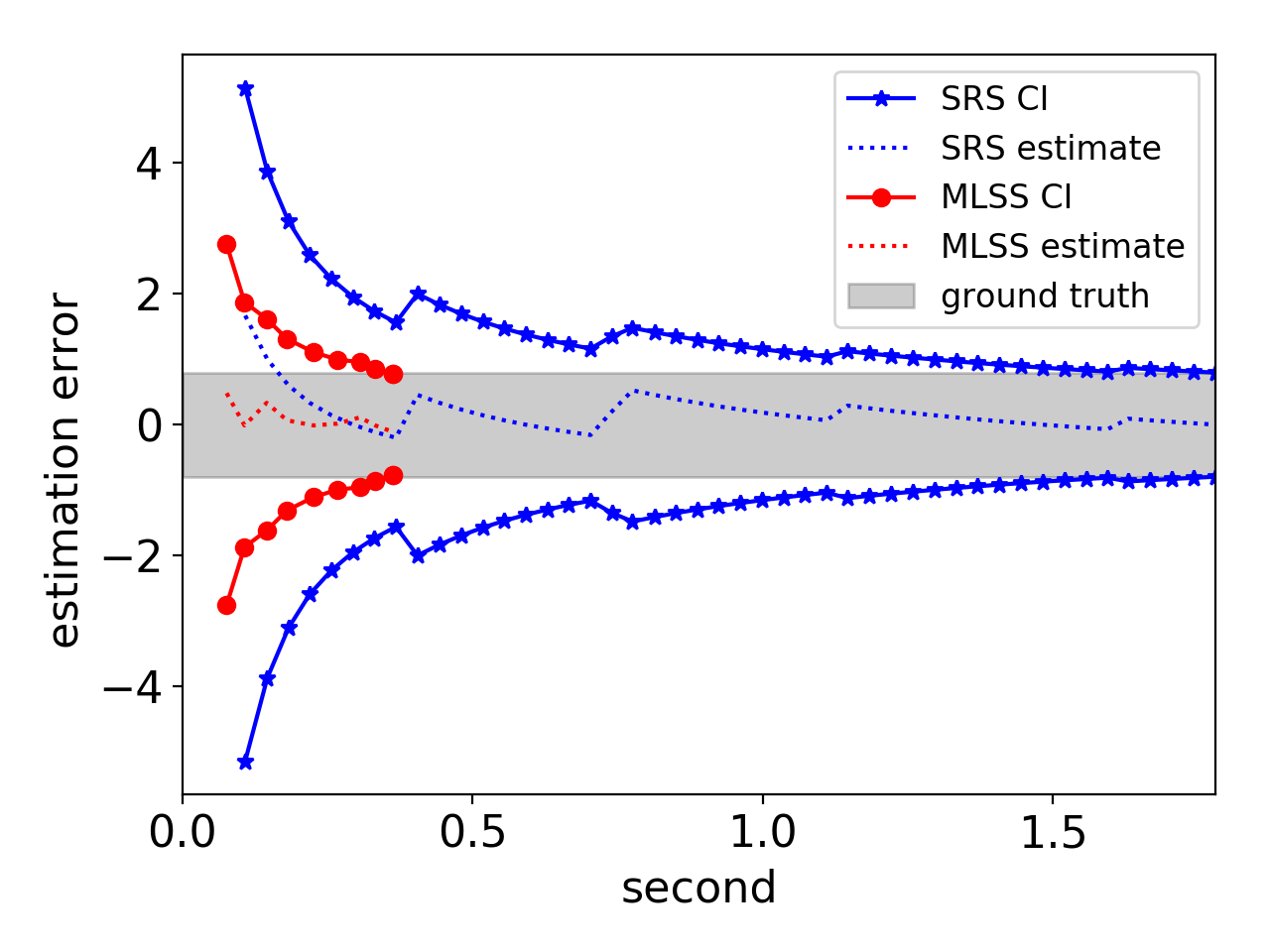}}
\subfloat[CPP Model, Tiny Query, RE]{\includegraphics[width=0.3\textwidth]{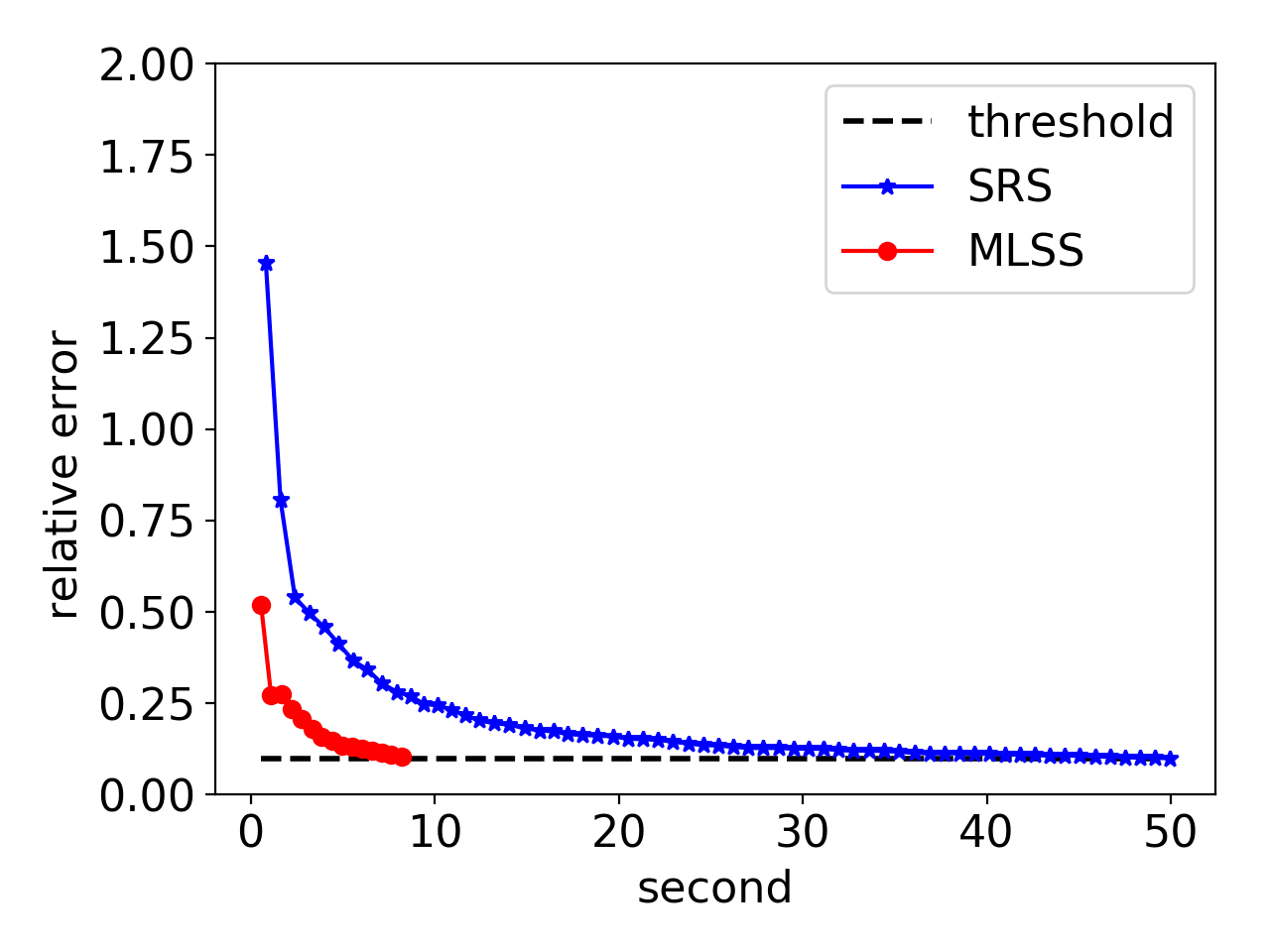}}
\subfloat[RNN Model, Tiny Query, RE]{\includegraphics[width=0.3\textwidth]{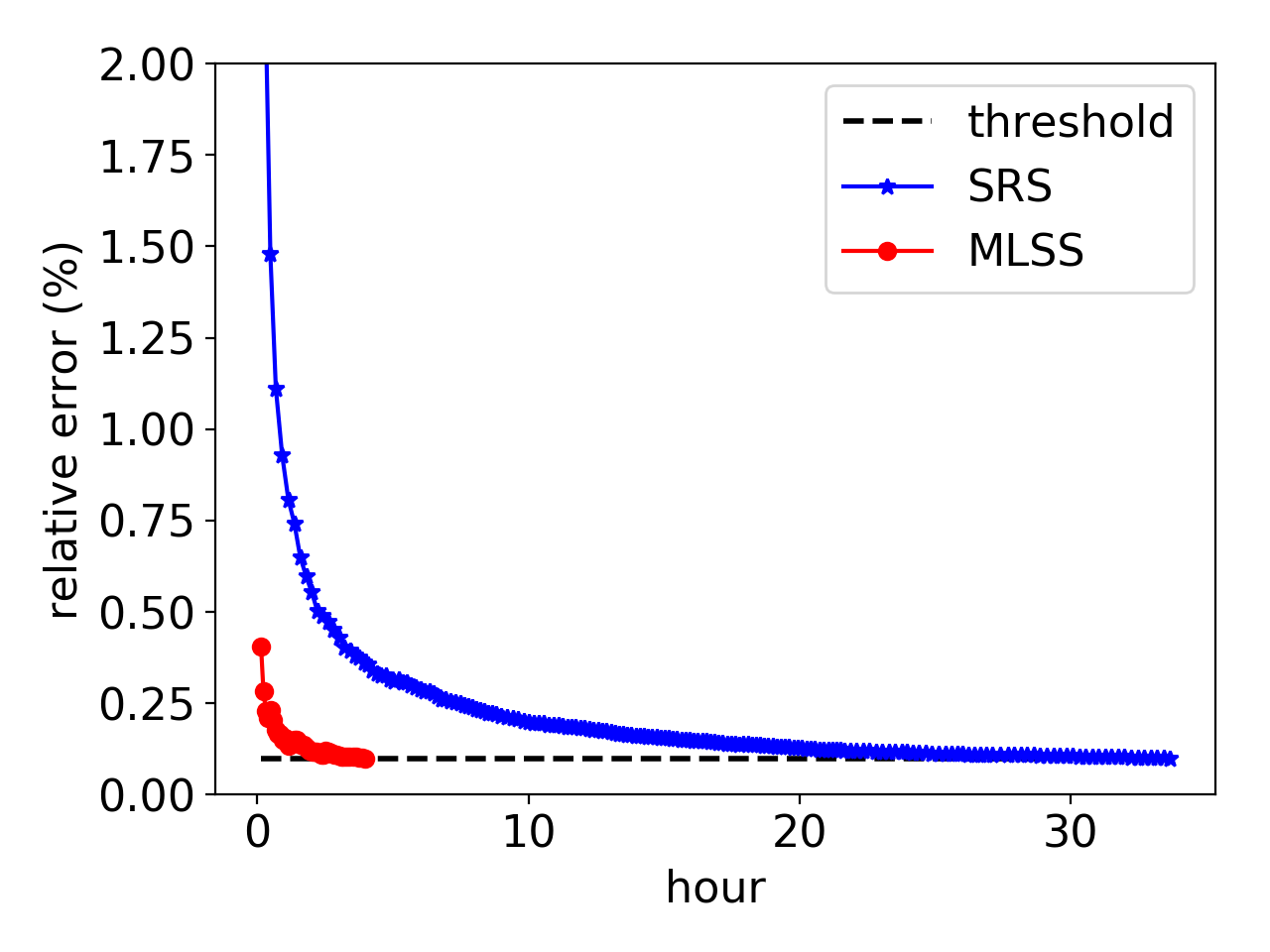}}
\caption{Query answer quality over time.}
\label{fig:ans-quality-over-time}
\end{figure*}

\subsection{MLSS vs. SRS}\label{sec:expr:mlss-vs-srs}
In this section, we comprehensively compare the performance of MLSS and SRS.
\ansa{
Note that under the settings of experiments in this particular section, ``level skipping'' will not occur, so g-MLSS is equivalent to s-MLSS.
Therefore, we do not distinguish between s-MLSS and g-MLSS in this section, and just use the term MLSS for simplicity to compare with the baseline.
In Section~\ref{sec:expr:g-MLSS}, we modify the processes to make them more volatile (so ``level skipping'' will happen) and further evaluate the performance of g-MLSS.
}
For each stochastic temporal model, we design four types of durability queries: Medium, Small, Tiny and Rare, denoting the quantity of the (estimated) true answer probability of the queries.
Detailed query parameters are summarized in Table~\ref{tab:query-setting}.
\mparagraph{Estimations and Overall Efficiency}
We first demonstrate the answer quality, i.e., unbiasedness, of MLSS.
For each model and for each type of query, we repeatedly run SRS and MLSS 100 times, respectively, and average the returned answers along with empirical standard deviations.
Results are summarized in Tables~\ref{tab:ans-queue} (Queue Model), \ref{tab:ans-cpp} (CPP Model) and, \ref{tab:ans-nn} (RNN Model).
As shown in these tables, the answers (hitting probability) returned by SRS and MLSS, on all types of queries and on all models, are essentially the same.
Even though they are not identical, the differences are well within the standard deviation.
This finding confirmed our analysis and proof in Section~\ref{sec:mlss} about MLSS's unbiased estimation.

Next, let us compare the query efficiency between SRS and MLSS.
We time the query until its answer (estimation) achieves certain quality target.
As shown in Figure~\ref{fig:queue-efficiency} (Queue Model) and Figure~\ref{fig:cpp-efficiency} (CPP Model),
MLSS generally runs significant faster than SRS (note the log scale on y-axis).
For Medium and Small queries, we can see a 40\% to 60\% query time reduction brought by MLSS.
For Tiny and Rare queries, MLSS runs 10x faster than SRS, without loss of answer quality.
As discussed earlier in Section~\ref{sec:mlss-vs-srs}, the main advantage of MLSS to SRS is the ability to focus and encourage simulations that move towards the target.
This property is especially helpful for those durability queries with lower probability, since MLSS can better distribute simulation efforts to promising paths hitting the target, instead of blindly wasting time on those failure paths (which would be a large portion of the total) as SRS did.
We observe similar query efficiency improvement on the more complex RNN model.
In Table~\ref{tab:ans-nn}, for Small and Tiny queries (which are more commonly asked in practice) on RNN model, there is a roughly 80\% to an order-of-magnitude query time reduction provided by MLSS.

Overall, MLSS clearly surpasses SRS across different models and on different types of commonly asked durability queries in practice, providing query speedup from 40\% up to an order of magnitude, without sacrificing answer quality.
\ansb{
It is also worth mentioning that MLSS is best suited for Tiny and Rare queries, and may not provide much benefit for larger queries. 
If the target is relatively easy to reach (which corresponds to large probability), the splitting behavior of MLSS would bring little benefit and may result in unnecessary overhead. Hence, in later sections, we would focus our discussions/evaluations more on Tiny and Rare queries, which are also the commonly asked durability queries in practice.
}

\mparagraph{Query Performance over Time}
To take a closer look at query performance comparison of MLSS and SRS, we monitor query answers and its quality (CI or RE) over time, and plot the convergence of estimations on single run of MLSS and SRS, respectively.
See Figure~\ref{fig:ans-quality-over-time} for details.
In Figure~\ref{fig:ans-quality-over-time}(1), we run a Small query on Queue model and use CI as estimation quality measure.
For better illustration, CI intervals are interpreted as percentage to the true probability such that it will be centered at 0.
The grey ribbon in the plot shows the desired region for a reliable estimate (true probability with 1\% CI).
Symmetric red lines and blue lines demonstrate how CIs of MLSS and SRS converge over time.
Red dotted line and blue dotted line are the estimate of MLSS and SRS over time.
It is clear that MLSS converges faster than SRS on estimation quality.
On the other hand, we can also see that the estimates (red dotted line and blue dotted line) from MLSS and SRS are always nicely contained by its corresponding CI, showing the statistical guarantees brought by CI.
We observe similar behaviors on CPP model (Figure~\ref{fig:ans-quality-over-time}(2)) and RNN model (Figure~\ref{fig:ans-quality-over-time}(3)).
Here we use run Tiny queries on these two models and use RE as quality measure.
Similarly, the time that MLSS needs for a reliable estimate (10\% RE, dashed line in the plot) is significantly shorter than that of SRS.
The fast convergence of MLSS's estimator further explains why MLSS can be notably efficient than SRS in general.
\subsection{Simple MLSS vs.  General MLSS}\label{sec:expr:g-MLSS}
\begin{figure}
    \centering
    \subfloat[Query time, Volatile CPP]{\includegraphics[width=0.24\textwidth]{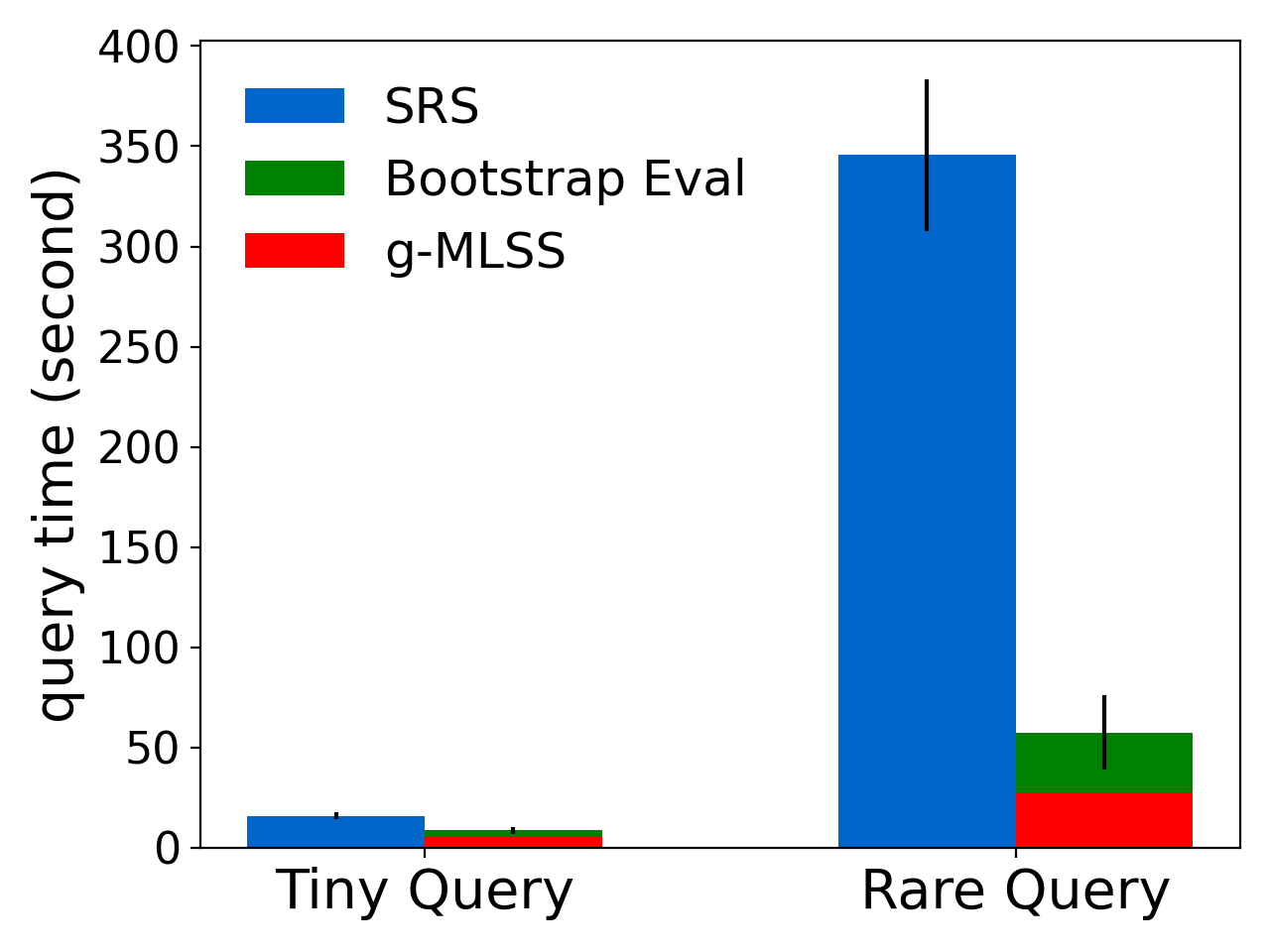}}
    \subfloat[Query time, Volatile Queue]{\includegraphics[width=0.24\textwidth]{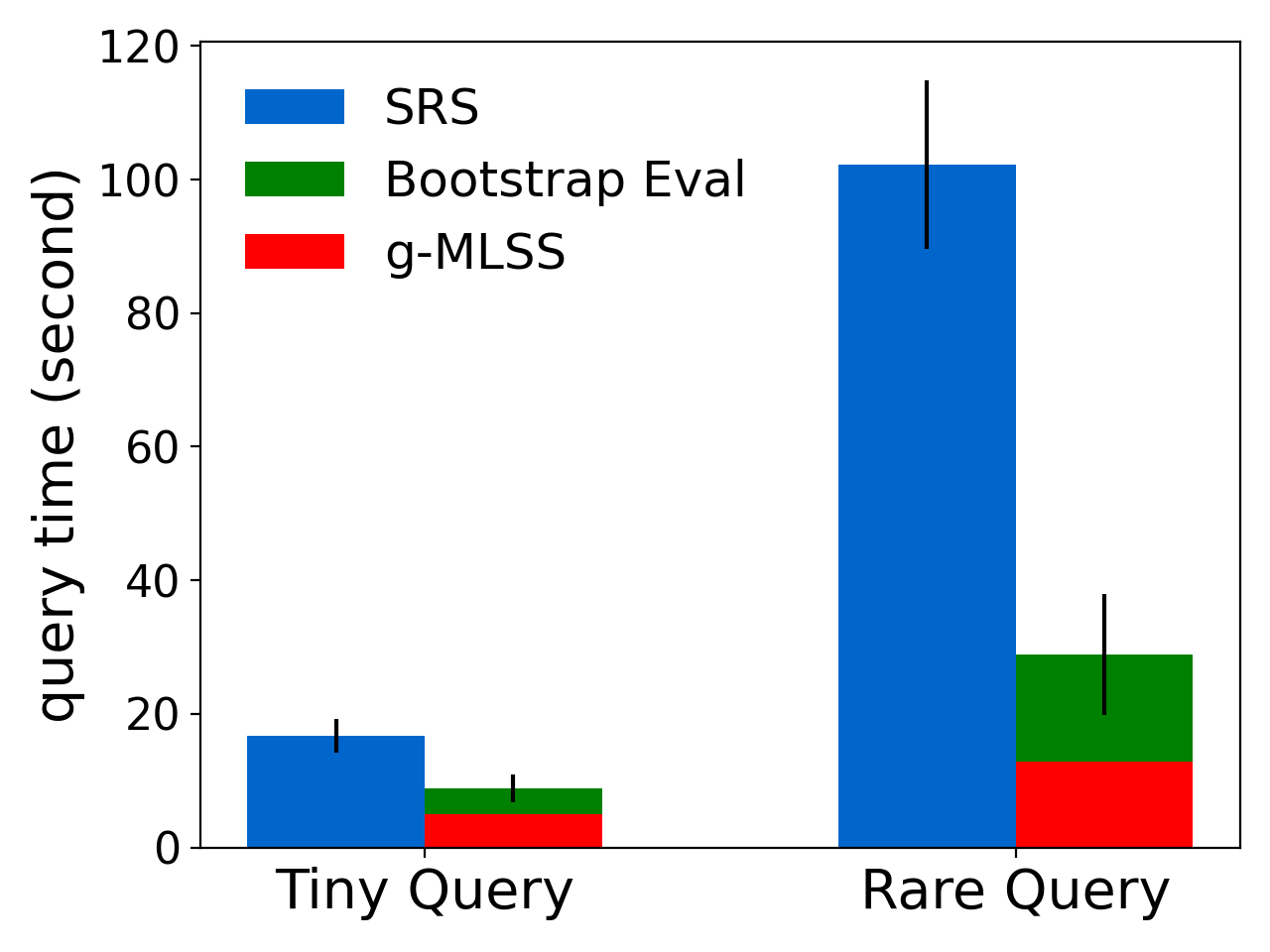}}
    \caption{g-MLSS query efficiency on models with volatile value changes.}
    \label{fig:cpp-g-mlss}
    \vspace{-1em}
\end{figure}

\begin{table}[t]
    \centering
    \caption{Performance comparison between s-MLSS and g-MLSS on temporal process with volatile values changes.}
    \label{tab:g-mlss}
    \resizebox{0.5\textwidth}{!}{
    \begin{threeparttable}
    \begin{tabular}{|c|c|c|c|c|}
         \cline{2-5}
         \multicolumn{1}{c|}{} & \multicolumn{2}{c|}{Volatile CPP} & \multicolumn{2}{c|}{Volatile Queue} \\\cline{2-5}
    \multicolumn{1}{c|}{} & \multicolumn{1}{c|}{\makecell{Tiny Query \\ ($s:500, \beta:700$)}} & \multicolumn{1}{c|}{\makecell{Rare Query \\ ($s:500, \beta:1000$)}} & \multicolumn{1}{c|}{\makecell{Tiny Query \\ ($s:500, \beta:65$)}}
    & \multicolumn{1}{c|}{\makecell{Rare Query \\ ($s:500, \beta:75$)}}\\\cline{1-5}
    SRS & 2.2\%$\pm$1.5\% & 0.1\%$\pm$0.2\% & 1.7\%$\pm$0.9\% & 0.3\%$\pm$0.26\%\\\hline
     s-MLSS & 1.1\%$\pm$0.8\% & 0.05\%$\pm$0.07\% & 1.2\%$\pm$0.5\% & 0.2\%$\pm$0.11\%\\\hline
     g-MLSS & 2.1\%$\pm$1.2\% & 0.09\%$\pm$0.1~\% & 1.7\%$\pm$0.5\% & 0.3\%$\pm$0.17\% \\\hline
    \end{tabular}
    \end{threeparttable}
    }
\end{table}
s-MLSS works well in practice if the underlying process satisfies the no level-skipping assumption, as demonstrated in previous sections.
\ansb{%
To show the limitation of s-MLSS and the generality of g-MLSS, we consider level skipping by experimenting with new temporal processes based on CPP model and Queue model with impulse value jumps between consecutive time instants.
More specifically, when $t>0.8s$ we introduce large value increase of (200 for CPP, and 5 for Queue) with small probabilities ($0.005$ for CPP and $0.2$ for Queue).
We refer to these two processes as Volatile CPP and Volatile Queue.

\mparagraph{Estimation} First, we test the unbiasedness of our approach.
we fix the simulation budget (i.e., 50000 invocations to the simulation procedure) and compare average estimations with empirical standard deviation based on estimates obtained from 100 independent runs.
Results are summarized in Table~\ref{tab:g-mlss}.
It is clear that, with the existence of level skipping, s-MLSS gives wrong estimates.
In contrast, g-MLSS still provides unbiased estimation by gracefully handling level-skipping paths, and it has higher precision (smaller standard deviation) compared to SRS under the same simulation budget.

\mparagraph{Overall Efficiency} Second, we evaluate the query efficiency of g-MLSS.
Figure~\ref{fig:cpp-g-mlss} shows the performance of g-MLSS on Volatile CPP and Volatile Queue.
Recall from our discussions in Section~\ref{sec:extension} that we do not have an analytical expression for g-MLSS variance; instead, we implement bootstrap sampling to empirically estimate it for evaluating the stopping condition.
Hence, the bootstrap evaluation time is also counted towards the total query time (shown in green in the plot).
Overall, as presented in  Figure~\ref{fig:cpp-g-mlss}, g-MLSS beats SRS by a large margin. Especially for Rare, we can see nearly 80\% improvement on both models.
Focusing on the breakdown of total query time, we can see that the bootstrap evaluation takes up a large portion (more than 50\%) of the query time.
Our currently implementation of bootstrapping is rather unoptimized; with more sophisticated implementations, we expect g-MLSS to still have plenty of room for further efficiency improvement.
}
\subsection{MLSS Optimization}\label{sec:expr:mlss-opt}
\begin{figure}[t]
    \centering
    \subfloat[Queue Model]{\includegraphics[width=0.24\textwidth]{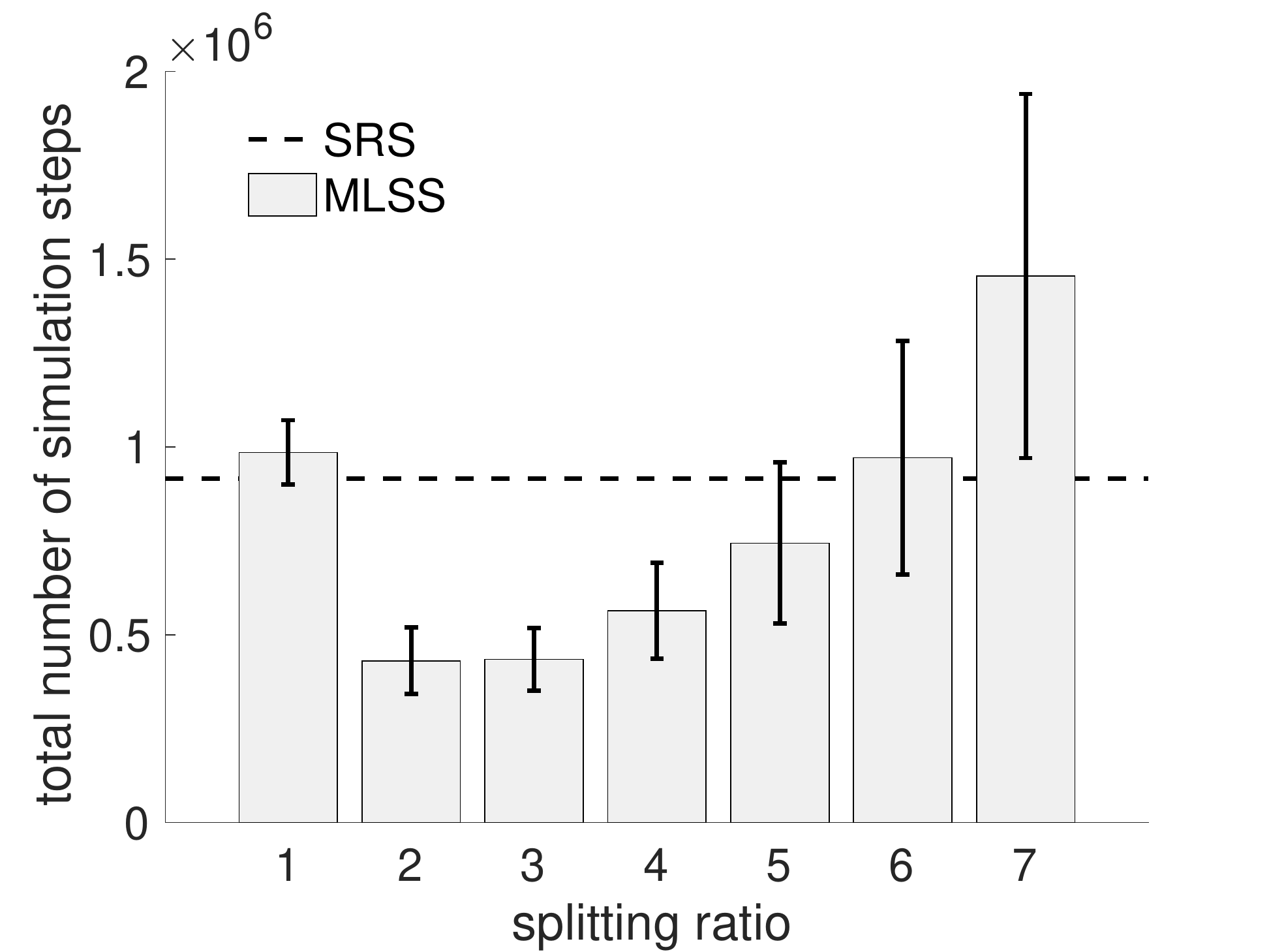}}
    \subfloat[CPP Model]{\includegraphics[width=0.24\textwidth]{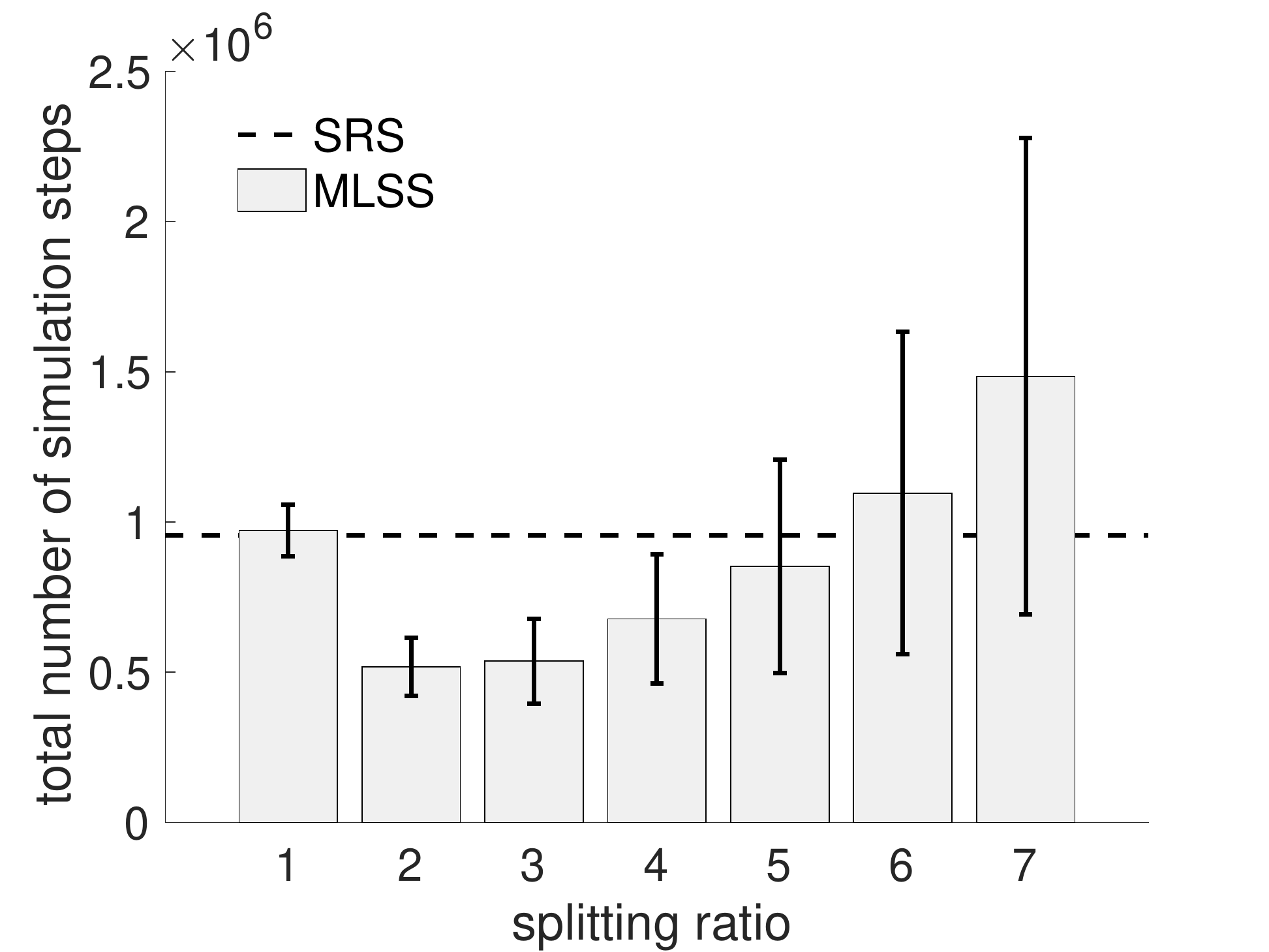}}
    \caption{Trade-off between splitting ratio and MLSS's overall efficiency on Small Query.}
    \label{fig:optimal-r}
\end{figure}
\begin{figure}[t]
    \centering
    \subfloat[Queue Model]{\includegraphics[width=0.24\textwidth]{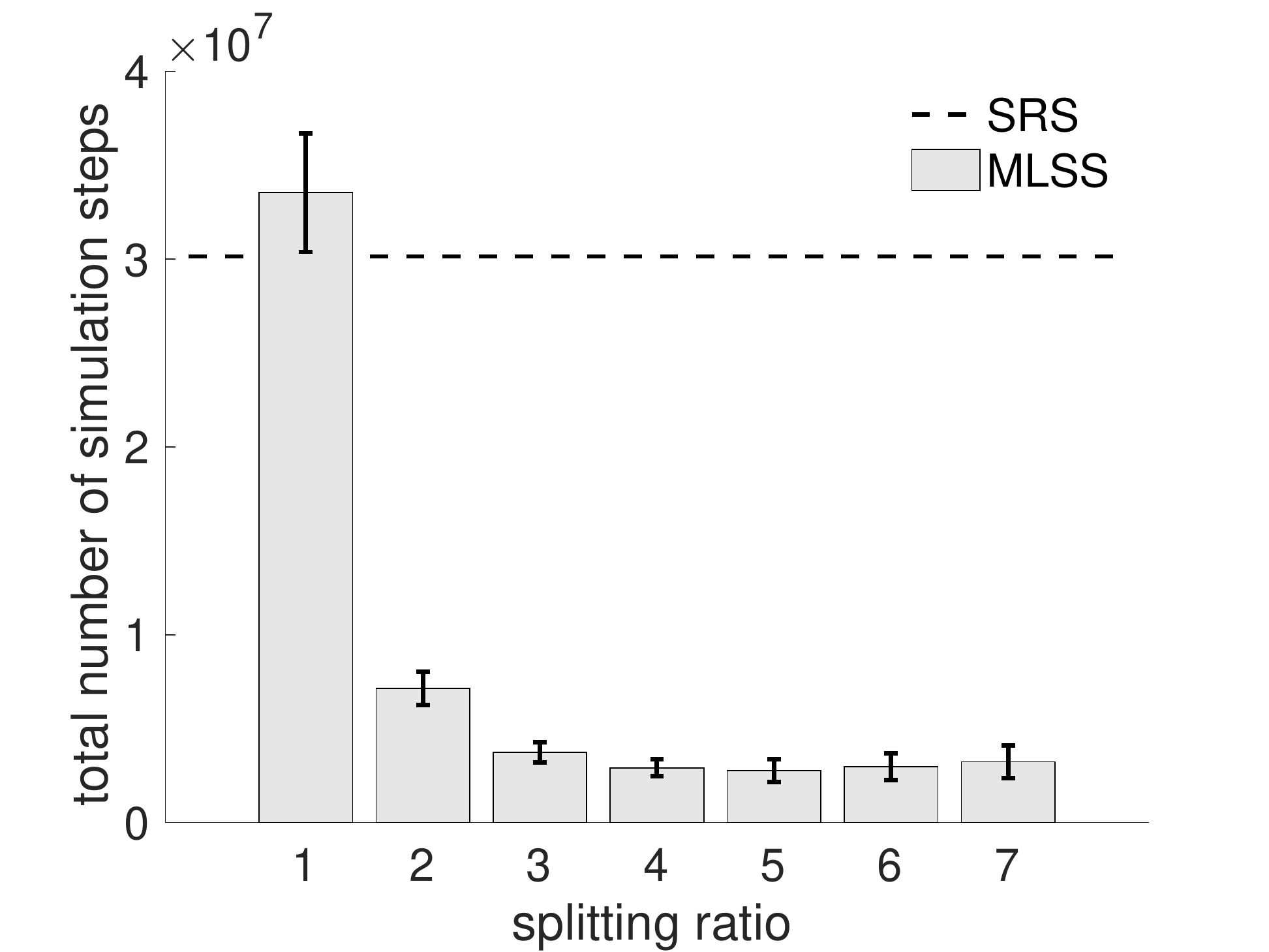}}
    \subfloat[CPP Model]{\includegraphics[width=0.24\textwidth]{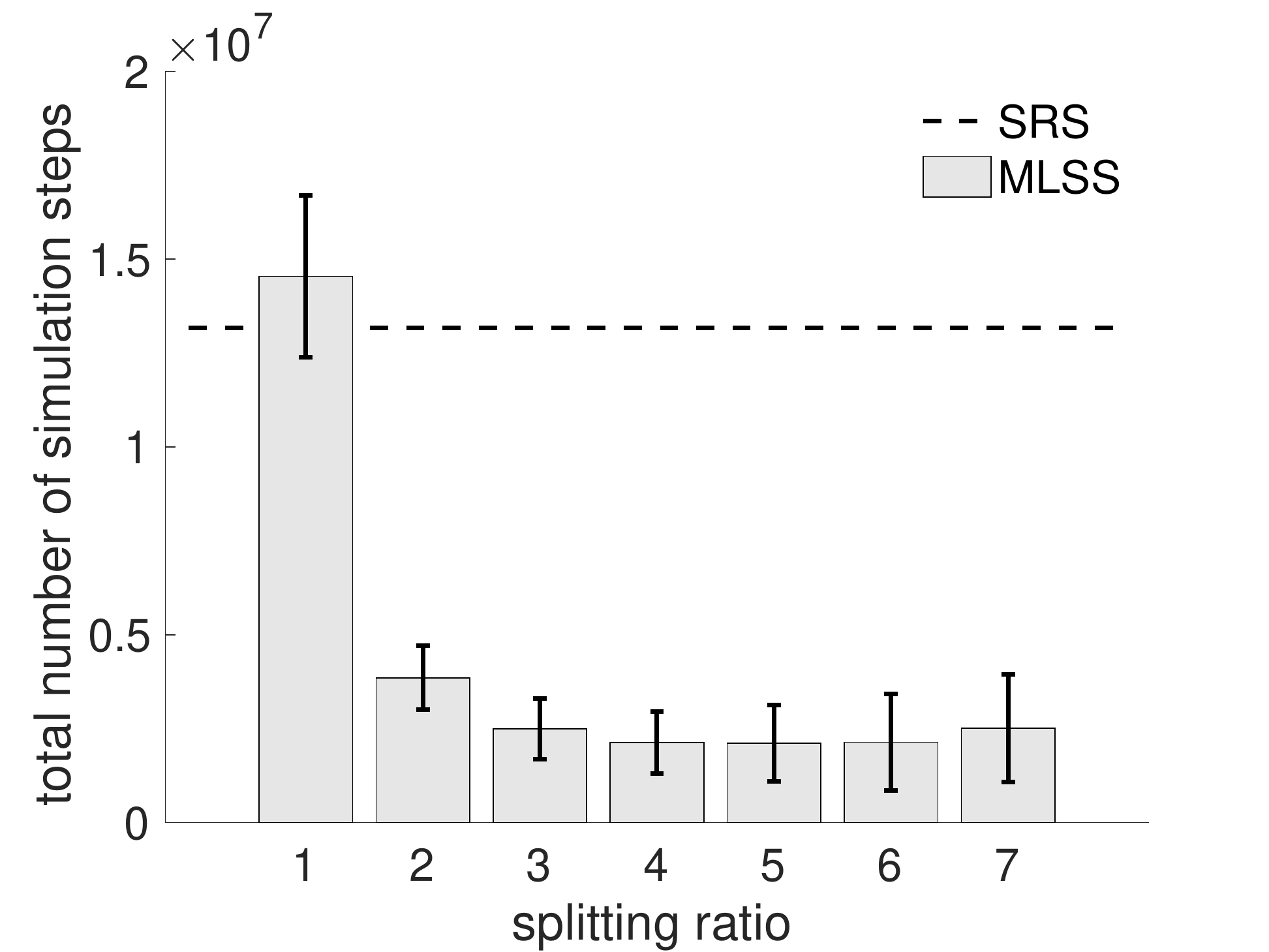}}
    \caption{Trade-off between splitting ratio and MLSS's overall efficiency on Tiny Query.}
    \label{fig:optimal-r-tiny}
\end{figure}
\begin{figure*}
    \centering
    \subfloat[Queue Model, Small]{\includegraphics[width=0.24\textwidth]{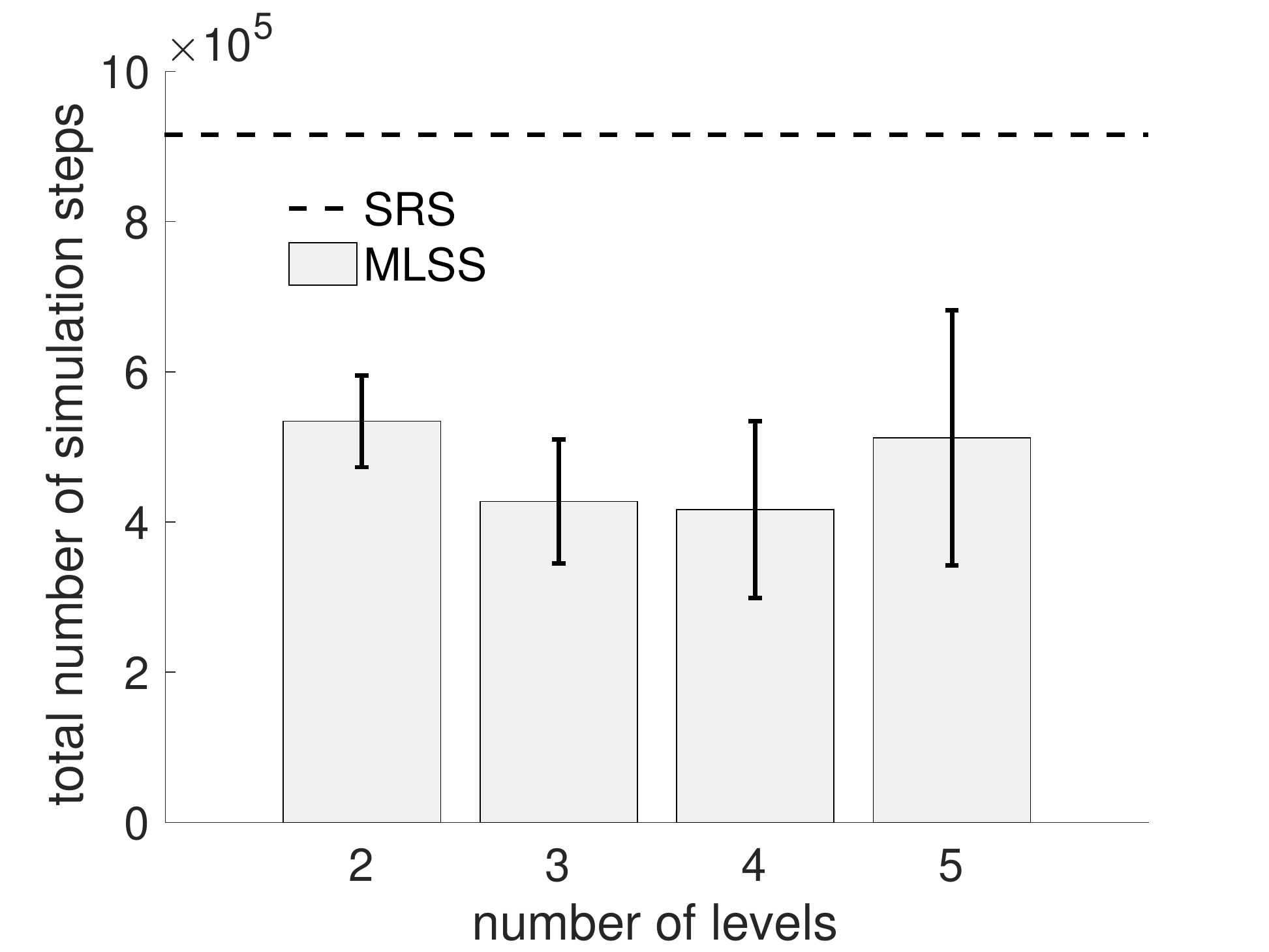}}
    \subfloat[CPP Model, Small]{\includegraphics[width=0.24\textwidth]{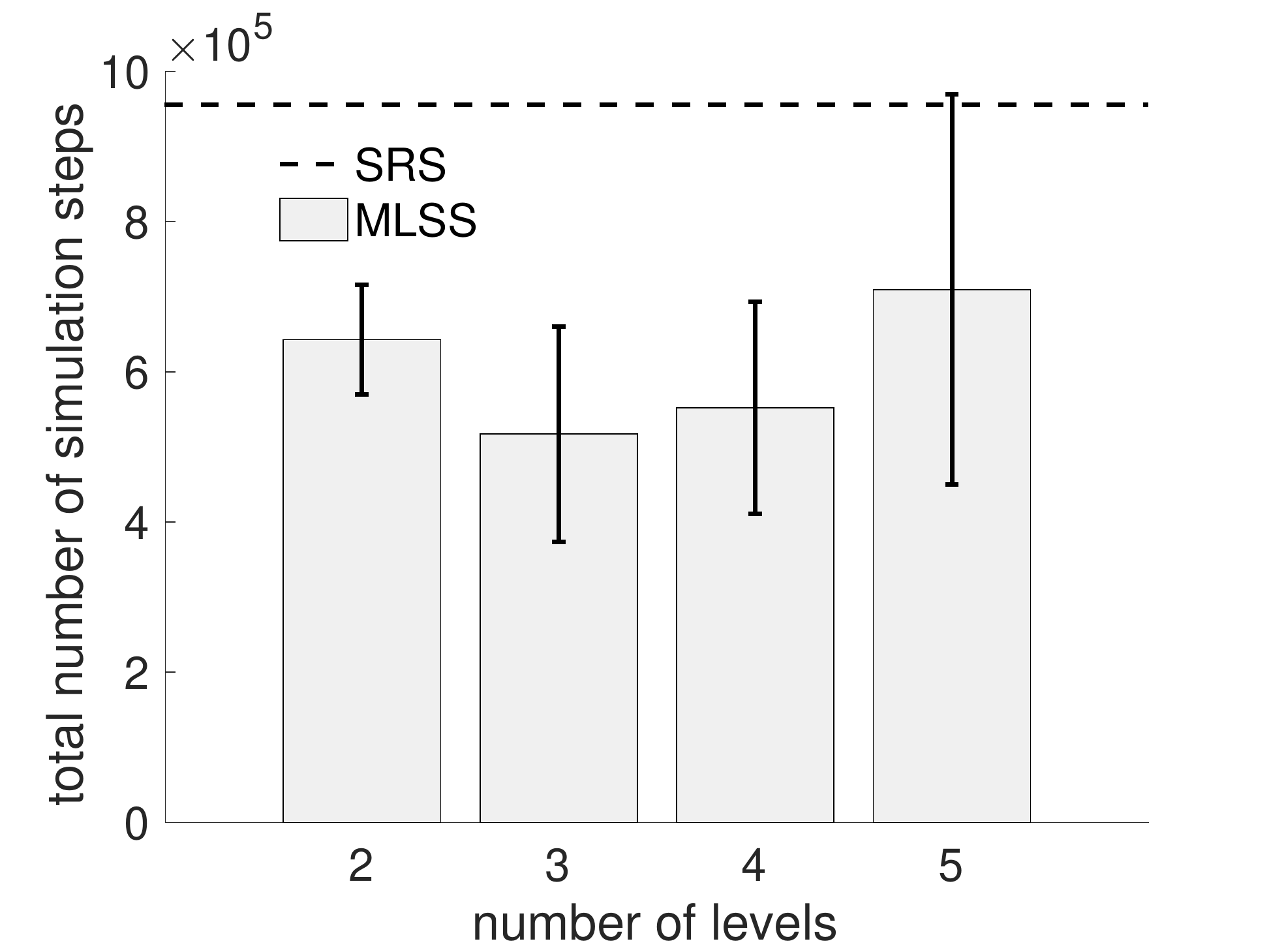}}
    \subfloat[Queue Model, Tiny]{\includegraphics[width=0.24\textwidth]{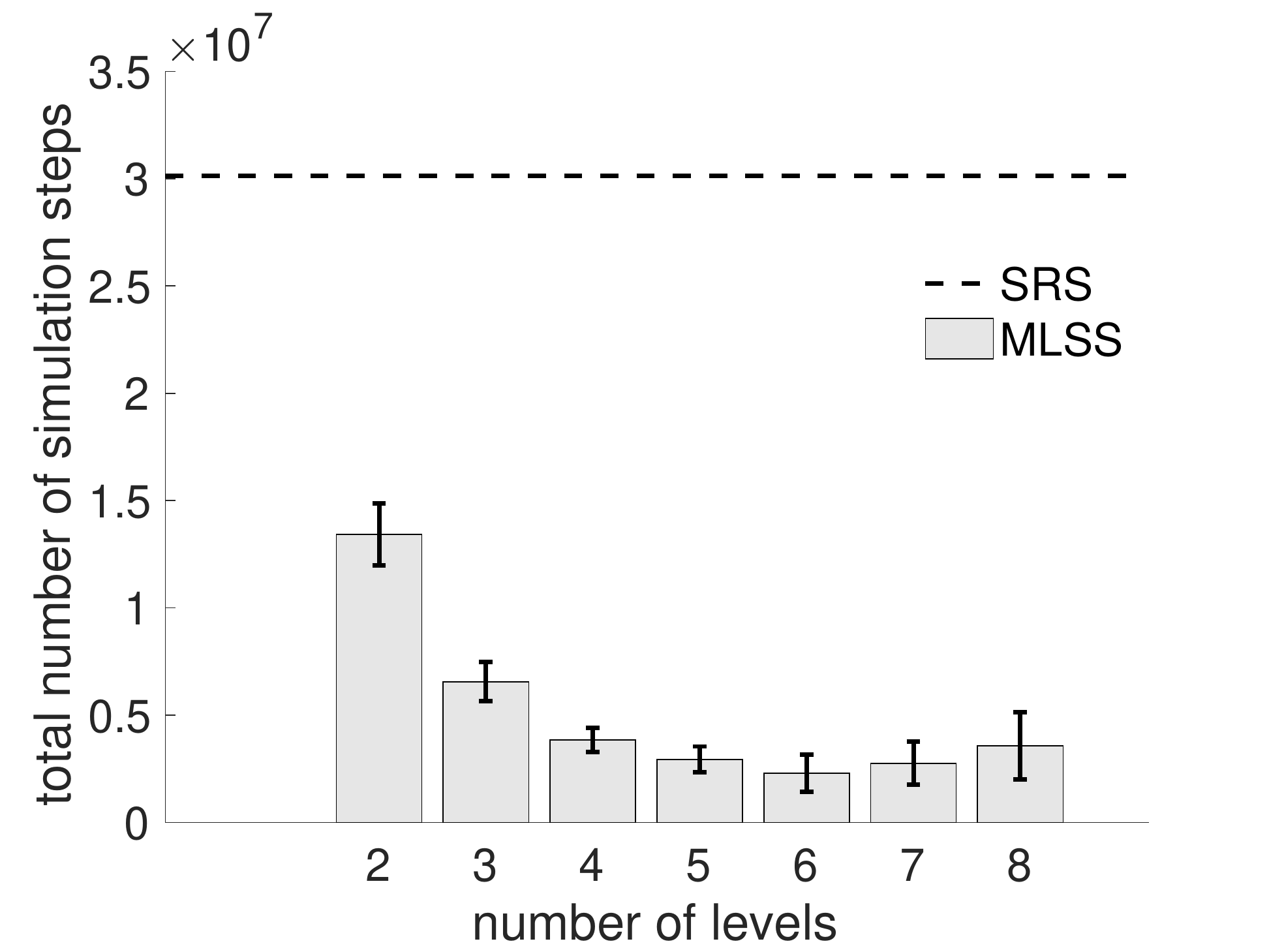}}
    \subfloat[CPP Model, Tiny]{\includegraphics[width=0.24\textwidth]{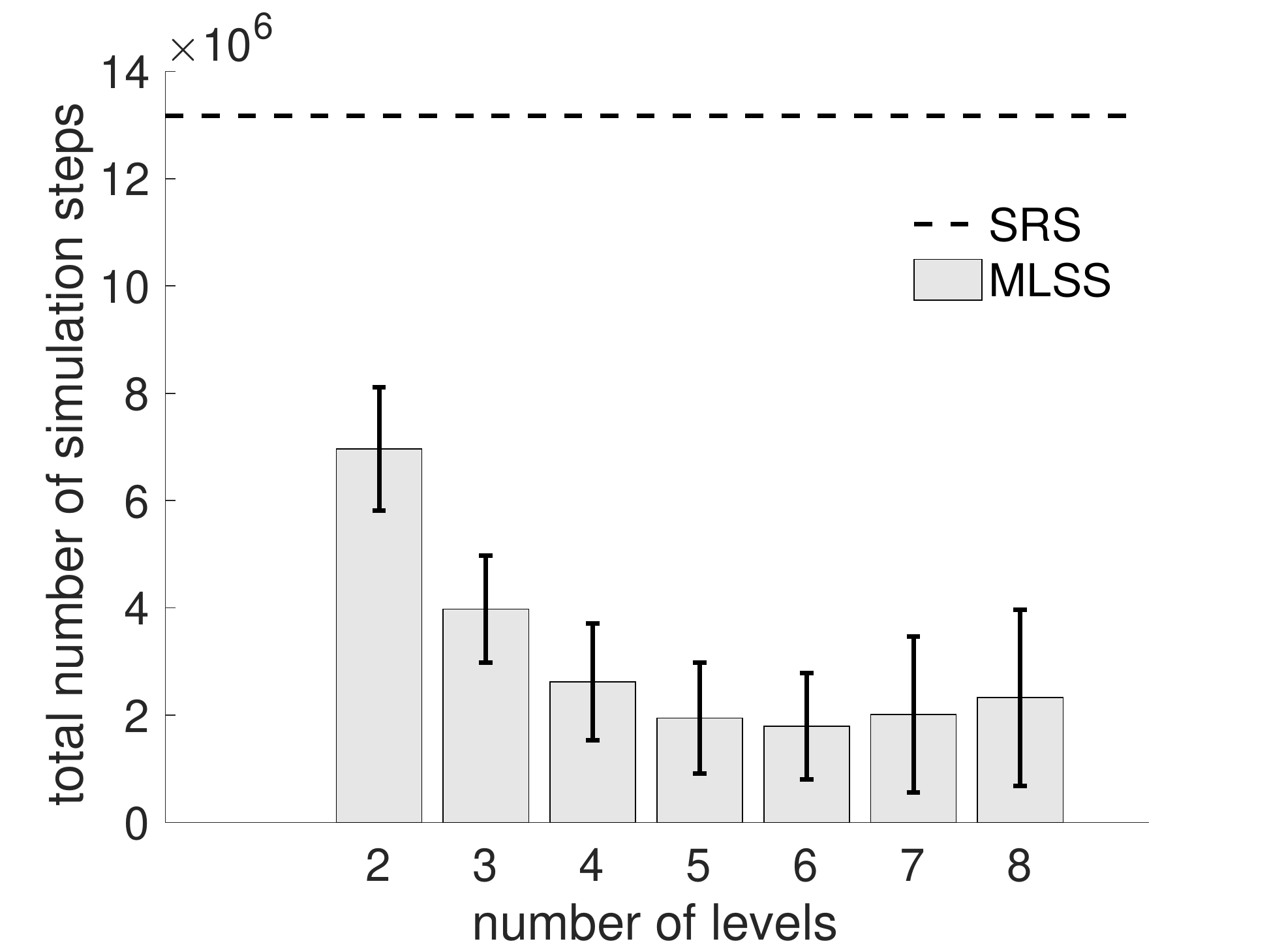}}
    \caption{Trade-off between number of levels and MLSS's overall efficiency on Small and Tiny Query.}\label{fig:optimal-levels}
\end{figure*}

\begin{figure*}
\begin{minipage}{0.69\textwidth}
\subfloat[Queue Model]{\includegraphics[width=0.33\textwidth]{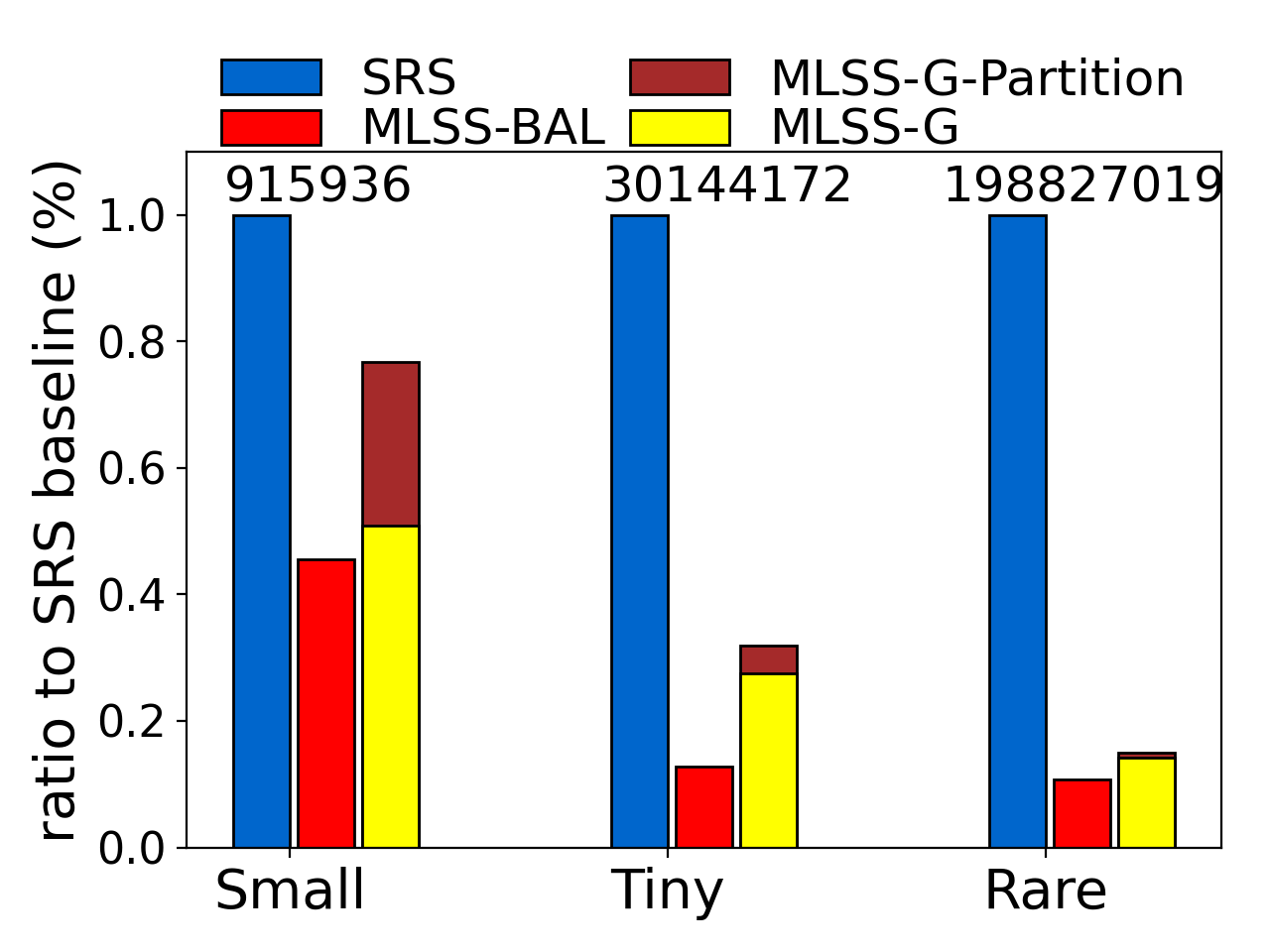}}
\subfloat[CPP Model]{\includegraphics[width=0.33\textwidth]{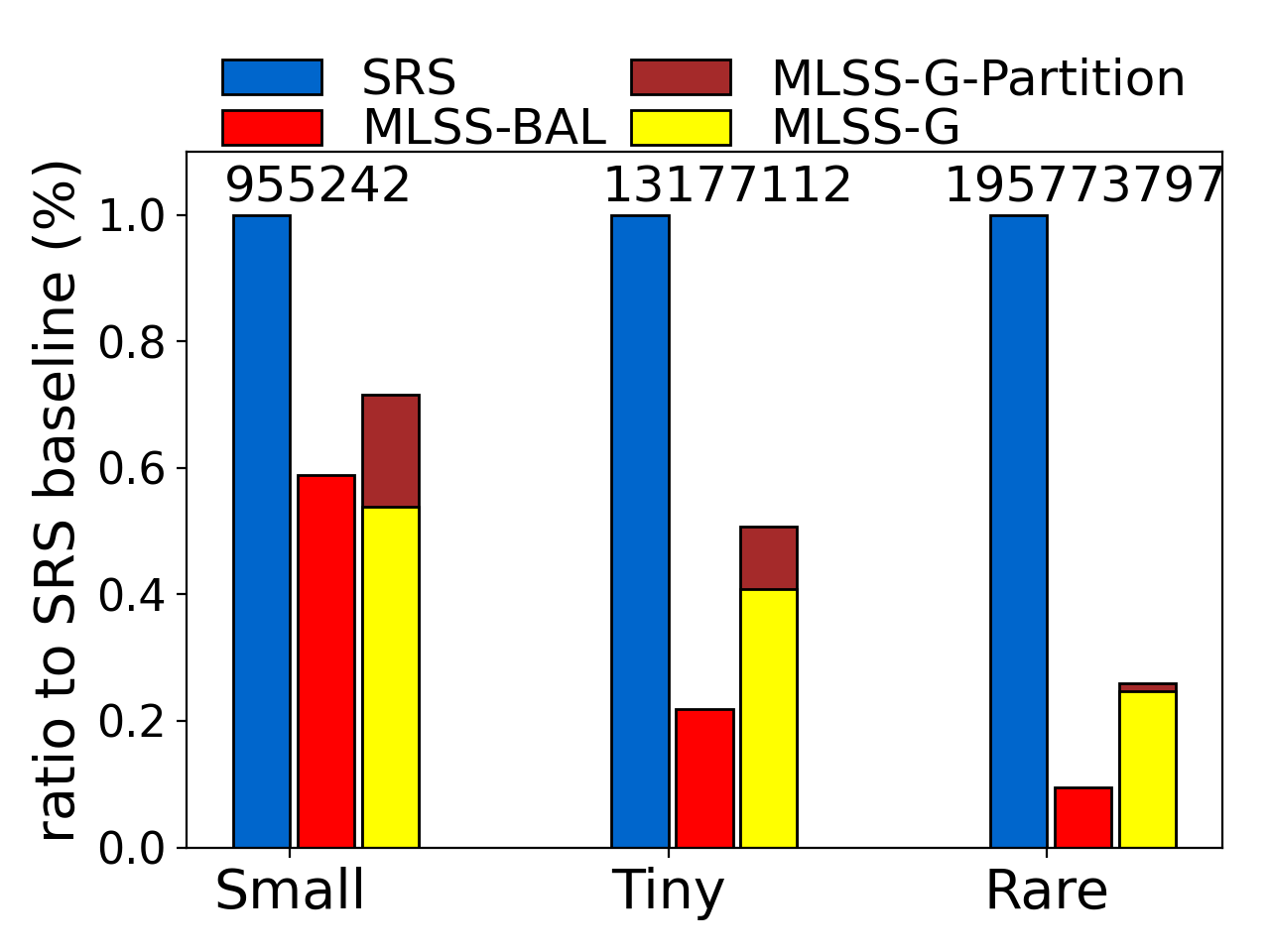}}
\subfloat[RNN Model (single run)]{\includegraphics[width=0.33\textwidth]{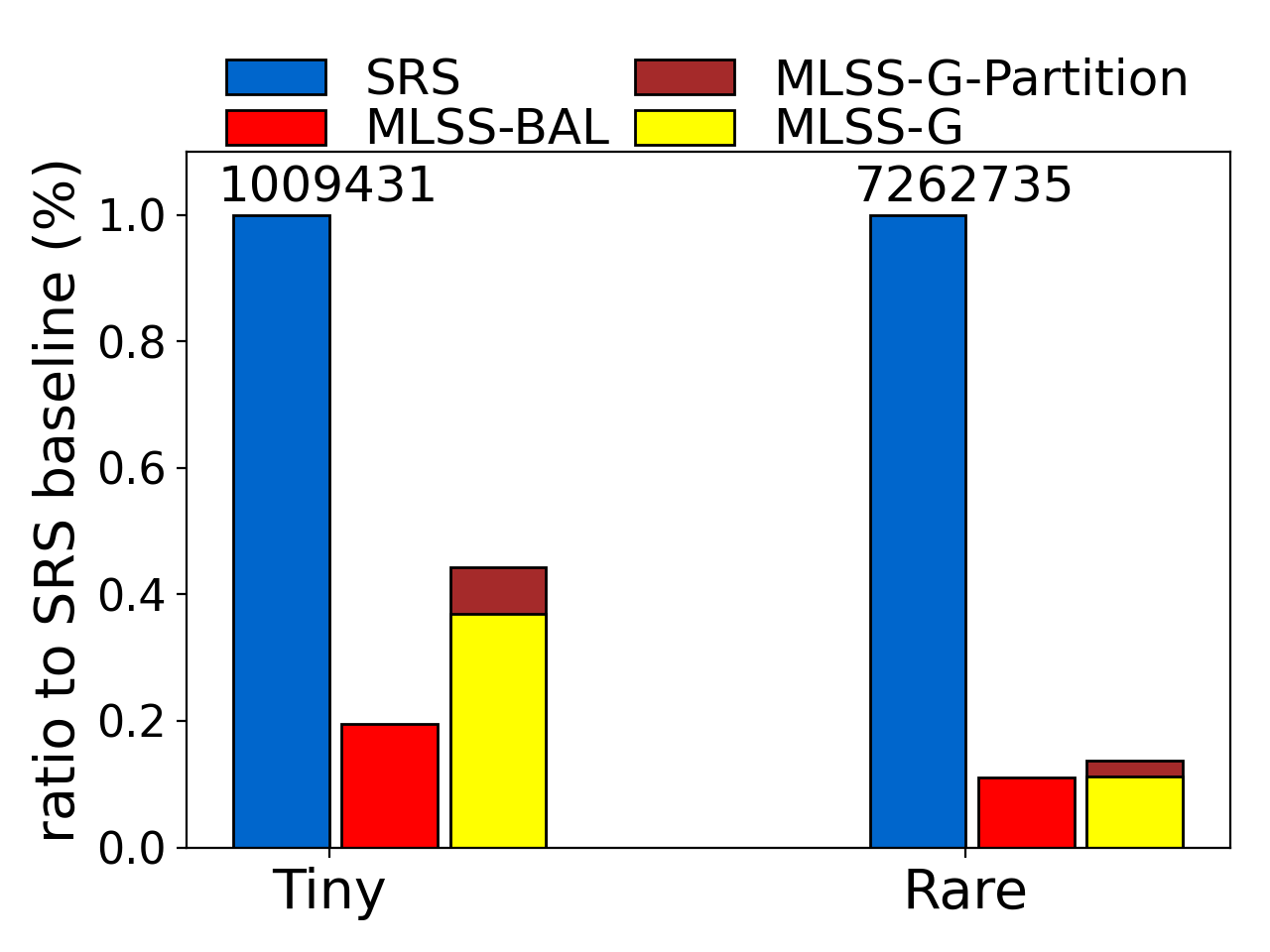}}
\caption{Efficiency of Greedy Level Partitions with s-MLSS.}\label{fig:greedy-efficiency}
\end{minipage}
\hfill
\begin{minipage}{0.3\textwidth}
    \captionsetup{type=table} 
    \centering
    \small
    \caption{Running times of MLSS and SRS on Queue/CPP model inside PostgreSQL. \mdseries Time usage in second.}
    \label{tab:mlss-dbms}
    \tabcolsep=0.1cm
    \begin{tabular}{|c|c|c|c|c|}
    \hline
        Queue Model & Medium & Small & Tiny & Rare \\\hline
        SRS & 6.6 & 2.4 & 146 & 1111 \\\hline
        MLSS & 6.1 & 1.3 & 23 & 79 \\\hline
        CPP Model & Medium & Small & Tiny & Rare \\\hline
        SRS & 10.2 & 3.7 & 112 & 3012 \\\hline
        MLSS & 8.8 & 2.5 & 27 & 173 \\\hline
    \end{tabular}
\end{minipage}
\end{figure*}

\begin{figure}[t]
    \centering
    \subfloat[Volatile Queue]{\includegraphics[width=0.24\textwidth]{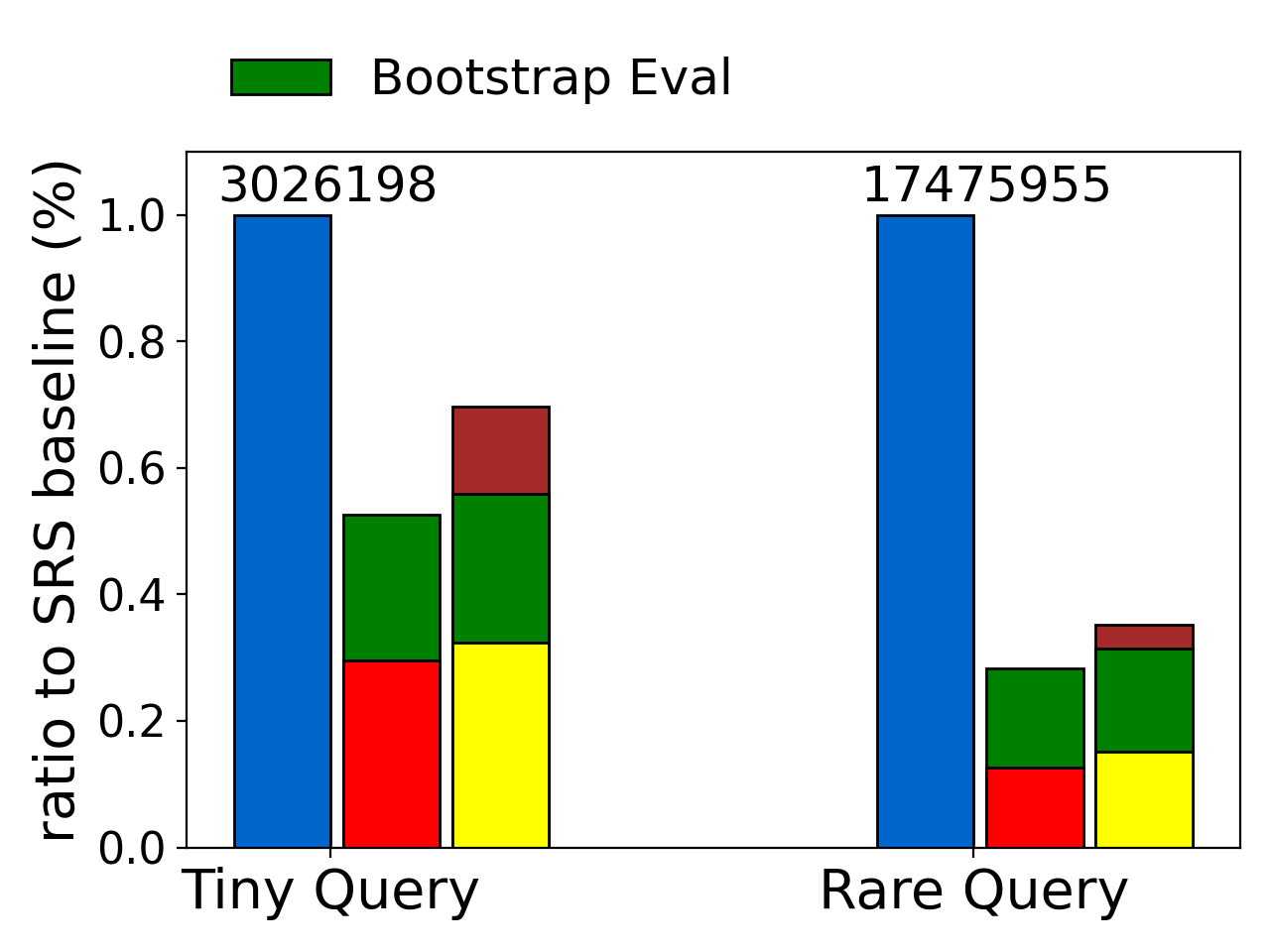}}
    \subfloat[Volatile CPP]{\includegraphics[width=0.24\textwidth]{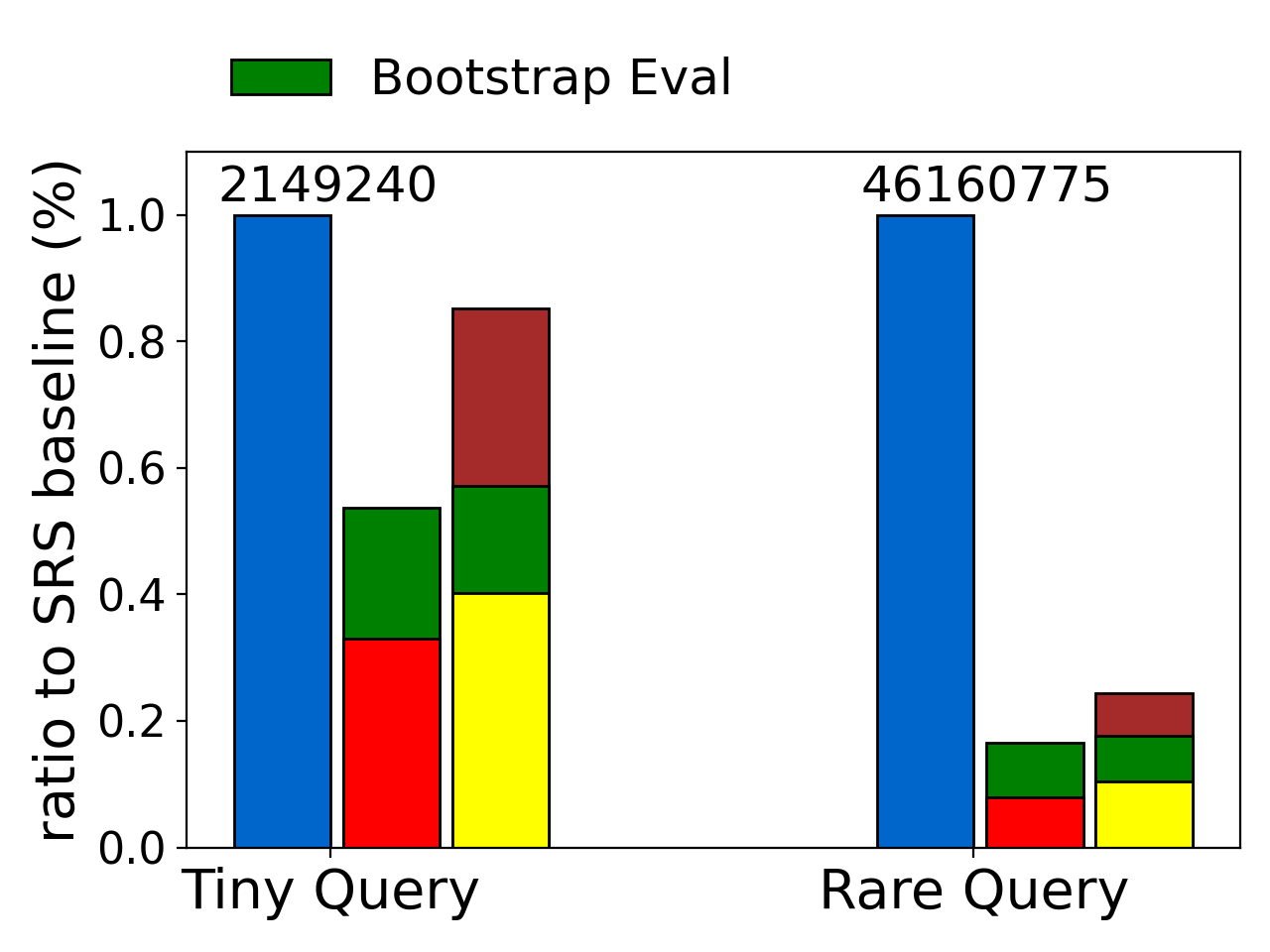}}
    \caption{Efficiency of Greedy Level Partitions on volatile temporal processes with g-MLSS.}
    \label{fig:g-mlss-greedy}
\end{figure}

In previous sections, we have shown the dominance of MLSS (both s-MLSS and g-MLSS) over SRS across a variety of models and query types.
We now focus more on MLSS itself, and investigate how sampling parameters of MLSS affect its overall efficiency and how to efficiently fine-tune MLSS in practice.
By default, we report average numbers over 100 trials, and standard deviations are shown as error bars on top of the plot.
\mparagraph{Optimal Splitting Ratios}
Figure~\ref{fig:optimal-r} shows a clear trade-off between splitting ratio and simulation efforts (to achieve reliable estimate; i.e., 1\% CI on Small Query and 10\% RE on Tiny Query).
To be fair, for splitting ratio from 1 to 7, we all use the ``balanced growth'' level partition strategy with four levels; that is, the crossing probabilities between consecutive levels are roughly the same.
Note that when splitting ratio is 1, MLSS is equivalent to SRS.
As shown in Figure~\ref{fig:optimal-r}, the first bar (splitting ratio is 1) matches with the SRS baseline (dashed line).
This finding further confirms the relationship between MLSS and SRS as proposed in Section~\ref{sec:mlss}.
It is not hard to understand the trade-off.
After all, a large splitting ratio directly leads to more splitted sample paths through the simulation process, not to mention that root paths are exponentially copied by the splitting ratio through multiple levels.
We can also see a clear difference of the optimal splitting ratio between Small Query and Tiny Query, where the latter prefers a larger value since it will potentially create more hits to the target (which is harder to achieve).
But interestingly, we can also observe that the optimal choice of splitting ratio (across different models and queries) seems fall in a narrow range around 3.
This is the reason why we fix splitting ratio as 3 as the default setting for MLSS.
Similar results can be found on CPP model as well.
\mparagraph{Optimal Number of Levels}
Figure~\ref{fig:optimal-levels} shows the relationship between the number of levels and the overall efficiency on Queue Model with Small and Tiny Query, respectively.
Here we fix splitting ratio as 3, and for different number of levels, we always adopt the ``balanced growth'' level partitions.
Again, SRS baseline is shown as dashed lines on top of the plots.
As these plots suggest, there is also a trade-off between the number of levels and the overall efficiency.
Recall the optimal theoretical result about ``balanced growth'' we introduced in Section~\ref{sec:optimal}.
More levels lead to smaller variance, but more levels also exponentially boost the splitting sample paths in each level of the simulations, which results in the aforementioned trade-off.
On the other hand, we observe that there does not seem to exist an universal optimal number of levels across queries.
For example, Small Query, Figure~\ref{fig:optimal-levels}(1), prefers fewer levels while Tiny Query, Figure~\ref{fig:optimal-levels}(3), requires 5 to 6 levels to achieve optimal performance.
This finding is consistent with the observations as we have on optimal splitting ratio from Figure~\ref{fig:optimal-r} that (compared to Small Query) Tiny Query requires more frequent target hits to achieve better performance.
\mparagraph{Greedy Level Partitions for s-MLSS}
After obtaining a better understanding of the factors that affect MLSS's overall efficiency,
finally, we consider the practical solution for efficiently fine-tuning MLSS based on our proposed greedy partition strategy introduced in Algorithm~\ref{algo:greedy}.
The baseline partition plans for comparison are the corresponding ``balanced-growth'' partition plans, which are obtained via manual tuning under the balanced-growth guideline and have the optimal number of levels.
In the following, we will refer to MLSS using such pre-tuned balanced-growth partition plans as MLSS-BAL; we do not charge the cost of manual tuning to running MLSS-BAL.
Figure~\ref{fig:greedy-efficiency} shows the effectiveness of the proposed greedy strategy on s-MLSS (in terms of the resulting overall running time to meet a given quality ).
For better visualization, we normalize all running times as ratios relative to the SRS baseline; hence in all plots, SRS is shown as the blue bars with ratio 1.
We also show the total number of simulation steps on top of each bar.
\ansb{
Red bars represent the running times of MLSS-BAL (recall that they do not include the cost of fine-tuning to find their balance-growth partition plans).
Yellow bars (MLSS-G) and brown bars (MLSS-G-Partition) together reflect the total running times of MLSS using the greedy algorithm,
with MLSS-G-Partition representing the search overhead of greedy algorithm.}
Overall, across all three models and different types of queries, the greedy algorithm is able to find \ansb{a partition plan comparable to the manually tuned ``balanced growth'' plan---the cost of MLSS-G is not so far away from MLSS-BAL,%
\footnote{\ansb{As can be seen in Figures~\ref{fig:greedy-efficiency}(2) and (3), sometimes our greedy strategy can discover a even better partition plan than MLSS-BAL.
This should not be surprising because the optimality of plans with balanced growth is based on certain assumptions (Section~\ref{sec:optimal}) that may not hold in practice.
Nonetheless, MLSS-BAL is a reasonable yardstick for comparison because the balanced growth guideline is, to the best of our knowledge, the only one that offers some theoretical guarantee of optimality.}}
and is still significantly lower than SRS with a 60\% to an order-of-magnitude improvement.
The search overhead (MLSS-G-Partition) is 10\% to 30\% of the total cost;
importantly, overhead seems lower for harder cases of Tiny and Rare, making greedy adaption an attractive approach in practice.}
\mparagraph{Greedy Level Partitions for g-MLSS}
\ansb{
We further test the greedy strategy on g-MLSS under volatile stochastic processes in Figure~\ref{fig:g-mlss-greedy}.
Again, we use (a rather unoptimized implementation of) bootstrapping to estimate the variance of g-MLSS in order to test the stopping condition;
that cost is shown as a green bar.
Overall, the total cost of g-MLSS with greedy adaptation (including greedy search overhead and bootstrapping overhead on top of simulation time) is lower than SRS in most cases acceptably close to MLSS-BAL, which has the benefit of pre-tuning.
Compared to the SRS baseline, our fully automated approach has a $\sim$20\% speedup on Tiny query and up to 80\% improvement on Rare query on both models.
}

In sum, considering that the greedy strategy does not need any information in advance and can automatically search for partition plans, it is a reasonable approach to try in practice if users do not have related knowledge of the model or the query.
\subsection{Implementations inside DBMS}\label{sec:dbms}
The database management system (DBMS) provides a single platform for not only data management, transformation, and querying, but also increasing machine learning support~\cite{boehm2019data}.
Predictive models, ranging from classic statistical models (i.e., Queue and CPP) to complex learning-based models, can be seamlessly encoded inside DBMS for data analytics; e.g., MCDB~\cite{jampani2008mcdb}. 
MLSS can also be straightforwardly integrated into a DBMS by implementing its sampler and estimator as stored procedures.
In this section, we move the query answering pipeline inside a DBMS (PostgreSQL), including both the predictive models and query processing algorithm.
More specifically, we use a database table for storing parameters of the procedure $\SimStep$ to allow step-by-step forward simulations and implement MLSS as stored procedure using Python Procedural Language.
We repeat our experiments as in Section~\ref{sec:expr:mlss-vs-srs} and report results in Table~\ref{tab:mlss-dbms}.
We see the advantage of MLSS over SRS as in earlier experiments; for example, we brought the running times of Rare queries from 0.3-0.8 hour required by SRS to under a few minutes.
This demonstrates sufficient promises towards an end-to-end ML lifecycle inside DBMS: data ETL (Extract, Transform and Load), building predictive models, and efficiently answering durability queries based on predictions for various data analytics.
Moreover, we can materialize sample paths generated from MLSS simulations as separate database tables, which can be further used for visualizations or other analysis.

\subsection{Summary of Experiments}
In conclusion, we first demonstrate (simple) MLSS's strong dominance over SRS across different stochastic processes and different types of queries.
In general, we observe a query time speedup from 50\% (Medium-to-Small queries) up to an order-of-magnitude (Tiny-to-Rare queries), without sacrificing answer quality.
The general MLSS extends simple MLSS to more general predictive models and continues to provide higher quality answer to durability queries (with 20\% to 80\% query time reduction) compared to SRS.
Next, we further inspect MLSS and investigate factors that affect its overall efficiency, and present a greedy strategy that frees users from time-consuming and tedious parameter optimizations.
With nearly no information needed in advance, the proposed greedy strategy automatically searches for near-optimal setting (according to the empirically evaluations introduced in Section~\ref{sec:plan-eval}) simply through simulations.
Our experimental results confirm the effectiveness and efficiency of the greedy algorithm --- near-optimal query efficiency with only 10\% to 30\% search overhead of the overall simulation efforts.
Nonetheless, the greedy strategy is an alternative remedy if users have no knowledge about the stochastic process or queries.
Domain knowledge of the underlying models would definitely be helpful to facilitate the parameter tuning process, but is beyond the scope of this paper.
Finally, we demonstrate the possibility of integrating the whole durability prediction query processing pipeline inside DBMS, and show the vision of a DBMS-based end-to-end ML lifecycle.
\section{Related Work}\label{sec:related-work}
The closest line of work to ours is query processing over probabilistic databases~\cite{dalvi2009probabilistic}: range search queries
~\cite{cheng2003evaluating,cheng2004efficient,tao2005indexing,tao2007range}, top-$k$ queries~\cite{ge2009top, hua2011ranking, hua2008efficiently, re2007efficient, soliman2007top, yi2008efficient}, join queries~\cite{cheng2006efficient, kimelfeld2007maximally} and skyline queries~\cite{pei2007probabilistic}.
But there is a fundamental difference between these previous studies and our problem.
In this paper, we consider query processing based on predictive models that predict future temporal data, where temporal dependence is not neglectable when modelling data uncertainty.
As a comparison, previous work on probabilistic databases mainly focuses on the static (snapshot) data, where data uncertainty is considered independently for individuals.

Another similar line of work is MCDB and its variants~\cite{jampani2008mcdb, arumugam2010mcdb, perez2010evaluation, MCMC}.
Unlike probabilistic databases, MCDB does not make strong assumptions about uncertainty independence, but generally embodies data uncertainty with user-defined variable generation (VG) functions.
The use of VG functions is analogous to the way that we handle uncertain temporal data with predictive models.
Moreover, MCDB's solutions are simulation-based too.
The only difference is that our work devise novel sampling procedure to improve sampling efficiency while MCDB focuses on making standard Monte Carlo simulations run faster inside database management system.
In~\cite{emrich2012querying}, authors used Markov Chains to present uncertain spatio-temporal data and studied how to answer probabilistic range queries.
However, their solutions are specific to Markov Chains and requires the transition probability matrix as a priori information.
In contrast, our techniques are generally applicable to a variety of predictive models, and are largely independent of the underlying model itself.

Regarding durability queries, there are several papers exploring the notions of durability on temporal data.
In~\cite{lee2009consistent, leong2010durable, wang2013durable, gao2018durable}, authors consider durability as a fraction of times (that satisfies certain conditions) over a (temporal) sequence of snapshot data, and answer queries to return the top $k$ objects with highest durability.
In~\cite{jiang2011prominent, zhang2014discovering, jiang2009online}, authors view durability as the length of time interval.
They proposed that, on the two-dimensional space coordinating by durability and data values, skyline queries can discover interesting insights or facts from temporal data that are robust and consistent.
Though in different forms, these papers studied durability on existing historical data, which is certain.
To the best of our knowledge, our work is among the first to extend the notion of durability into the future, where data can only be probabilistic.

Sampling-based techniques and algorithms play an increasingly important role in the era of big data, ranging from data cleaning~\cite{sample-clean, marchant2017search}, integration~\cite{li2016wander} and evaluation~\cite{gao2019efficient}, to approximate query processing~\cite{chaudhuri2017approximate, jampani2008mcdb} and visualizations~\cite{sample-visual}.
Some of the work, e.g., \cite{marchant2017search, li2016wander}, share the same idea as ours---going beyond uniform sampling and improving sampling efficiency by properly granting (problem-specific) importance to positive samples.
The goal of this line of work is to reduce the total number of samples required without hurting the answer quality.
However, in many real-life applications, the actual cost of assessing selected samples (especially that involves manual work, i.e., labeling and annotation) might not be uniform.
Thus, the more accurate cost measurement of sampling-based solution in such application scenarios should be the actual cost (time or money) observed from practice, instead of just the number of samples.
Based on this consideration, some cost-aware sampling schemes~\cite{gao2019efficient} are proposed to practically alleviate the pain of expensive manual work.
Another direction is algorithm design for sampling-based solutions.
For example, MCDB~\cite{jampani2008mcdb} introduced the concept of tuple-bundle computations to make sampling-based query procedure run faster inside a database management system.
In~\cite{sample-visual}, based on online sampling-scheme, authors proposed an optimal incremental visualization algorithm to support rapid and error-free decision making.

Finally, our work also has a deep connection to two classic problems in statistic community: first hitting time~\cite{redner2001guide} and rare event simulation~\cite{bucklew2013introduction}.
In statistics, first hitting time (also known as first passage time or survival analysis) is an important feature of stochastic or random process, denoting the amount of time required for a process (starting from an initial state) to reach the threshold for the first time.
As mentioned in previous sections, it has a wide applications in very diverse domains~\cite{fauchald2003using, shiryaev1999essentials}.
Rare event simulation is the scenario that the probability of the event is low, say, order of $10^{-3}$ or less. 
In such cases, the standard Monte Carlo approach would fail to provide reliable estimate in an efficient manner.
Importance sampling~\cite{glynn1989importance} and splitting-based sampling~\cite{garvels2000splitting} are two popular variance reduction techniques for rare event simulations.
In this paper, we propose to apply Multi-Level Splitting Sampling, which is based on splitting-based sampling, as a query processing technique to efficiently answer durability prediction queries.

\section*{Acknowledgements}
This work was supported by NSF grants IIS-1718398, IIS-1814493, CCF-2007556, and a grant from the Knight Foundation.
Any opinions, findings, and conclusions or recommendations expressed in this publication are those of the author(s) and do not necessarily reflect the views of the funding agencies.

\newpage
\appendix
\section{Full Proof of Proposition~\ref{lemma:mlss-unbias}}\label{appendix:proof}

\begin{proof}
Consider any level $L_i$, let $S_i$ denote the set of entrance states of all possible paths that enter $L_i$, and $S_i^{\langle j \rangle}$ be a sample entrance state of $S_i$.
Then, for each $S_i^{\langle j \rangle}$, we can define a Bernoulli random variable variable $X_{i+1}(S_i^{\langle j \rangle}) \sim Bernoulli(\Delta_i^{\langle j \rangle})$ with probability $\Delta_i^{\langle j \rangle}$, denoting Bernoulli trials of a path (whether it reaches the next level $L_{i+1}$ starting from state $S_i^{\langle j \rangle}$ in $L_i$).
Note that $\Delta_{i}$ is a random variable with respect to the hitting probability from level $L_i$ to $L_{i+1}$, and $\Delta_i^{\langle j \rangle}$ is a sample value (of $\Delta_i$) based on entrance state $S_i^{\langle j \rangle}$.
An important observation is
\begin{equation}\label{eq:proof-1}
    \E[\Delta_{i}] = p_{i+1};
\end{equation}
that is, the expectation of success probability (of entering $L_{i+1}$) over all entrance states in $L_i$ equals $p_{i+1}$, which is the probability that a path enters $L_{i+1}$ conditioning on its entrance to $L_i$.

Now assume that during MLSS procedure, we have $N_i$ entrance states in $L_i$: $S_i^{\langle 1 \rangle}, S_i^{\langle 2 \rangle}, \dots, S_i^{\langle N_i \rangle}$.
For each of them we split and simulate $r$ independent copies and watch whether they hit the next level $L_{i+1}$.
Recall that the observations of each $S_i^{\langle j \rangle}$ is a Bernoulli variable with probability $\Delta_i^{\langle j \rangle}$, thus the number of hits to $L_{i+1}$ contributed by state $S_i^{\langle j \rangle}$ follows a \emph{binomial} distribution as $B(r, \Delta_i^{\langle j \rangle})$. 
Combining $N_i$ states together, we have the following formula of counter $N_{i+1}$:
\begin{equation}\label{eq:proof-2}
    N_{i+1} = \sum_{j=1}^{N_i}\sum_{k=1}^{r} X_{i+1}(S_i^{\langle j \rangle}) \sim \sum_{j=1}^{N_i} B(r, \Delta_i^{\langle j \rangle}).
\end{equation}
Since $N_{i+1}$ conditions on $N_i$ and the sampled entrance states, we further have

\begin{align}\label{eq:proof-3}
\begin{split}
    \E[N_{i+1}\mid N_i] & =  \E\bigg[\E[N_{i+1}\mid N_i, S_i^{\langle 1 \rangle}, S_i^{\langle 2 \rangle}, \dots, S_i^{\langle N_i \rangle}]\bigg] \\ &~~~(\text{law of total expectation})\\
    & = \E\bigg[\E[\sum_{j=1}^{N_i} B(r, \Delta_i^{\langle j \rangle})\mid N_i, S_i^{\langle 1 \rangle}, S_i^{\langle 2 \rangle}, \dots, S_i^{\langle N_i \rangle}]\bigg] \\
    &~~~(\text{by Eq(\ref{eq:proof-2})})\\
    &=\E[\sum_{j=1}^{N_i} r \cdot \Delta_i^{\langle j \rangle}] = N_ir\E[\Delta_{i}]\\ 
    &= N_i r p_{i+1}. ~~~\text{(by Eq(\ref{eq:proof-1}))}
\end{split}
\end{align}
Applying this to $\hat{p}_{i+1}$ results in 

\begin{equation}
    \E[\hat{p}_{i+1} \mid N_i] = \E[\frac{N_{i+1}}{rN_i} \mid N_i] = \frac{\E[N_{i+1}\mid N_i]}{rN_i} = \frac{N_i r p_{i+1}}{rN_i} = p_{i+1}
\end{equation}
Finally, unconditioning on $N_i$ by law of total expectation, 
\begin{equation}
    \E[\hat{p}_{i+1}] = \E\bigg[\E[\hat{p}_{i+1} \mid N_i]\bigg] = \E[p_{i+1}] = p_{i+1}.
\end{equation}

Next, we prove that $\E[\hat{p}_1 \hat{p}_2 \cdots \hat{p}_m] = \E[p_1 p_2 \cdots p_m]$ by induction on $m$.
First, we know that at the starting level all root paths are independently simulated, thus it is obvious that $\hat{p}_1$ is an unbiased estimator for $p_1$; that is, $\E[\hat{p}_1] = p_1$.
Then, let us assume that $\E[\hat{p}_1 \hat{p}_2 \cdots \hat{p}_{m-1}] = p_1 p_2 \cdots p_{m-1}$.
Hence, we have
\begin{align}\label{eq:proof-induction}
\begin{split}
    \E[\hat{\tau}_{mlss}] &= \E[\hat{p}_1 \hat{p}_2 \cdots \hat{p}_m]\\
    &= \E\bigg[\hat{p}_1 \hat{p}_2 \cdots \hat{p}_{m-1}\E[\hat{p}_m \mid N_m]\bigg] ~~~\text{(law of total expectation)}\\
    & = \E[\hat{p}_1 \hat{p}_2 \cdots \hat{p}_{m-1} p_m] ~~~\text{(by Eq(\ref{eq:proof-3}))} \\
    & = \E[\hat{p}_1 \hat{p}_2 \cdots \hat{p}_{m-1}]p_m\\
    & = p_1 p_2 \cdots p_m = \tau, ~~~\text{(by assumption)}
\end{split}
\end{align}
which finishes the induction and proves that $\hat{\tau}_{mlss}$ is an unbiased estimation of $\tau$.
\end{proof}

\balance
\bibliographystyle{ACM-Reference-Format}
\bibliography{ref.bib}

\end{document}